\documentclass[longauth]{aa} 
\usepackage{graphicx}
\usepackage{tabularx}
\usepackage{multirow}
\usepackage[colorlinks=true, linkcolor=blue, citecolor=blue]{hyperref}
\usepackage{gensymb}
\usepackage{color}
\usepackage{txfonts}
\usepackage{natbib}
\usepackage{amsmath}
\usepackage{ulem}
\usepackage[flushleft]{threeparttable}
\usepackage[dvipsnames]{xcolor}
\usepackage{rotating}
\usepackage{CJKutf8}
\usepackage{placeins}
\usepackage{afterpage}
\usepackage{stfloats}

\newcommand{\nii}{[N\,\textsc{ii}]}
\newcommand{\cii}{[C\,\textsc{ii}]}
\newcommand{\ci}{[C\,\textsc{i}]}

\newcommand{\ha}{H$\alpha$}

\begin{document} 

\titlerunning{Dormant Galaxies at $z > 3$ with ALMA}
\authorrunning{Xiao \& Qi et al.} 
   \title{Sleeping or Dead? Probing the Quenching Mechanism of JWST Dormant Galaxies at $z > 3$ with ALMA}

\author{Mengyuan Xiao\inst{\ref{unige}}\thanks{E-mail: mengyuan.xiao@unige.ch}
\and Songnan Qi\inst{\ref{unige}} 
\and Pascal A. Oesch\inst{\ref{unige},\ref{dawn},\ref{NBI}}
\and Longji Bing\inst{\ref{Sussex}}
\and Miroslava Dessauges-Zavadsky\inst{\ref{unige}}
\and Andreas L. Faisst\inst{\ref{Caltech}}
\and Mahsa Sanati\inst{\ref{unige}} 
\and Rashmi Gottumukkala\inst{\ref{dawn},\ref{NBI}}
\and Gabriel Brammer\inst{\ref{dawn},\ref{NBI}} 
\and David Elbaz\inst{\ref{cea}}
\and Robert Feldmann\inst{\ref{Zurich}}
\and Yoshinobu Fudamoto\inst{\ref{chiba}}
\and Lucas C. Kimmig\inst{\ref{nottingham}}
\and Benjamin Magnelli\inst{\ref{cea}}
\and Rui Marques-Chaves\inst{\ref{unige}} 
\and Anurag Amol Sawarkar \inst{\ref{unige}} 
\and Sune Toft\inst{\ref{dawn},\ref{NBI}}
\and Francesco Valentino\inst{\ref{dawn},\ref{DTU}}
\and Christina C. Williams\inst{\ref{NSF}}
}

\institute{Department of Astronomy, University of Geneva, Chemin Pegasi 51, 1290 Versoix, Switzerland\label{unige}
\and Cosmic Dawn Center (DAWN), Denmark \label{dawn}
\and  Niels Bohr Institute, University of Copenhagen, Jagtvej 128, K\o benhavn N, DK-2200, Denmark \label{NBI}
\and Astronomy Centre, University of Sussex, Falmer, Brighton BN1 9QH, UK  \label{Sussex}
\and Caltech/IPAC, 1200 E. California Blvd. Pasadena, CA 91125, USA \label{Caltech}
\and Universit{\'e} Paris-Saclay, Universit{\'e} Paris Cit{\'e}, CEA, CNRS, AIM, 91191, Gif-sur-Yvette, France \label{cea}
\and Department of Astrophysics, University of Zurich, Zurich CH-8057, Switzerland\label{Zurich}
\and Center for Frontier Science, Chiba University, 1-33 Yayoi-cho, Inage-ku, Chiba 263-8522, Japan \label{chiba}
\and University of Nottingham, School of Physics and Astronomy, Nottingham NG7 2RD, UK \label{nottingham}
\and DTU Space, Technical University of Denmark, Elektrovej 327, DK-2800 Kgs. Lyngby, Denmark\label{DTU}
\and NSF National Optical-Infrared Astronomy Research Laboratory, 950 North Cherry Avenue, Tucson, AZ 85719, USA \label{NSF}
}
\date{Received xxx; accepted xxx}

  \abstract
 {
JWST has revealed a population of ``dormant'' galaxies at $z>3$ with strong Balmer breaks and weak emission lines, whose star formation appears to have declined only recently. Whether these galaxies are undergoing a lasting transition toward quiescence or merely experiencing a temporary suppression of star formation remains unclear. We investigate their molecular gas content using archival ALMA observations of \cii~158\,$\mu$m and CO rotational transitions. Starting from all publicly available JWST/NIRSpec prism observations and using joint spectrophotometric fitting, we construct a parent sample at $z_{\rm spec}>3$ and $\log(M_{\star}/M_\odot)>9.5$ in the ALMA-accessible COSMOS, UDS, and GOODS-South fields. We select dormant galaxies based on their Balmer-break strength, H$\alpha$ equivalent width, and dust attenuation, together with a comparison sample of quiescent galaxies selected using an evolving specific star-formation rate criterion. Our final sample comprises 7 dormant galaxies and 15 quiescent galaxies with ALMA constraints on their molecular gas content (mostly 3$\sigma$ upper limits). We find that dormant galaxies exhibit systematically lower star formation rates averaged over the past 10\,Myr ($\mathrm{SFR}_{10\,\mathrm{Myr}}$) than over the past 100\,Myr ($\mathrm{SFR}_{100\,\mathrm{Myr}}$), supporting a recent rapid decline in their star formation. 
For the $\sim70\%$ of the sample with sufficiently constraining ALMA observations, their molecular gas fractions lie at least $\sim0.3-2$\,dex below those expected for main-sequence galaxies. Moreover, the molecular gas fractions of dormant galaxies show no systematic difference from those of quiescent galaxies in our sample.  
This indicates that dormant galaxies have already reached a gas-poor state shortly after quenching and favors a scenario in which the rapid shutdown of star formation is associated with depletion or removal of the molecular gas reservoir, rather than primarily with suppressed star formation efficiency while retaining a substantial molecular gas reservoir. Our results suggest that, if no substantial gas replenishment occurs in the future, dormant galaxies are more consistent with becoming ``\textit{dead}'' than with being temporarily ``\textit{sleeping}''.
 }

   \keywords{galaxies: high-redshift -- 
   galaxies: evolution -- 
   galaxies:  star-formation -- 
   submillimetre: galaxies}
   \maketitle
   
%

\section{Introduction}
Understanding how galaxies cease forming stars is a central question in galaxy evolution \citep[e.g., the recent review by][]{Whitaker2026_REVIEW}. The existence of massive quiescent galaxies at increasingly high redshifts \citep[e.g.,][]{Schreiber2018_JekyllHyde,Valentino2023, Carnall2023Natur, Glazebrook2024,deGraaff2025, Weibel2025, Xiao2026_massivesample} indicates that both the build-up of stellar mass and the subsequent shutdown of star formation can occur rapidly in the early Universe \citep[e.g.,][]{Boylan-Kolchin2023, Lovell2023, Xiao2024,Xiao2025_spiral, Lagache2026,Bing2026, Faisst2026}. However, observations of fully quiescent galaxies provide relatively indirect constraints on the physical processes responsible for quenching, since these systems are generally observed after their star formation has already been suppressed (for tens to hundreds of Myr). Identifying galaxies during, or immediately after, the shutdown of star formation therefore offers a more direct opportunity to constrain how quenching proceeds.

Recent JWST/NIRSpec observations have revealed a population of ``dormant'' galaxies at high redshift that appear to be caught close to the onset of quenching \citep[e.g.,][]{Strait2023, Looser2024Natur, Looser2025, Baker2025_miniquench, Baker2026_twominiquench, Trussler2025, Covelo-Paz2026, Khullar2026}. These galaxies exhibit strong Balmer breaks and weak nebular emission lines, indicative of evolved stellar populations together with little ongoing star formation. Their inferred star-formation histories further suggest a recent decline in the star-formation activity, which can be more apparent on short ($\sim10-30$ Myr) than on longer ($\sim100$ Myr) timescales.
One possible interpretation of such short-lived declines is bursty star formation, which can produce temporary low-SFR phases ($\sim20-40$\,Myr) within burst-quench cycles at high redshift \citep[e.g.,][]{Sun2023_burstysf, Dome2024, Looser2024Natur, Faisst2025, McClymont2025}.
Therefore, dormant galaxies provide a view of systems shortly after a rapid decline in star formation, before they necessarily satisfy conventional quiescent-galaxy selections based on longer-timescale SFR indicators \citep[e.g., total UV and IR luminosities, see][]{Kennicutt2012}. Compared with more evolved quiescent galaxies, dormant galaxies may thus better preserve direct information about the physical conditions associated with the shutdown of star formation, making them particularly valuable laboratories for studying quenching mechanisms.

An important open question is whether this dormant phase represents the immediate route toward truly quiescent galaxies (\textit{dead}'' if no further gas is accreted) or, instead, a temporary suppression of star formation before renewed growth (\textit{sleeping}''). This question is closely connected to the physical origin of quenching. Broadly, the rapid decline in star formation can be driven by two scenarios: 1) depletion or removal of the molecular gas reservoir \citep[e.g., through rapid gas consumption during starbursts, feedback-driven outflows, or stripping;][]{Dekel2006, Tacchella2016b, Wang2026_redpotato}; or 2) suppression of the star-formation efficiency (SFE) while a substantial gas reservoir remains \citep[e.g., through morphological quenching or dynamical stabilization;][]{Martig2009, Xiao2018, Gensior2020}. These two scenarios make fundamentally different predictions for the molecular gas content ($M_{\rm mol}$) and molecular gas fraction ($\mu_{\rm mol}=M_{\rm mol}/M_{\star}$), and have different implications for their subsequent evolution. In the former case, galaxies are expected to be gas-poor shortly after quenching, with molecular gas fractions approaching those of quiescent galaxies. Without subsequent gas replenishment, they are likely to continue evolving toward complete quiescence. In the latter case, galaxies may retain substantial molecular gas reservoirs despite their low SFRs, potentially providing the fuel for renewed star formation at later times. Direct measurements of the molecular gas reservoir therefore provide a critical means of distinguishing between these two scenarios.

Previous studies of the molecular gas content of quiescent galaxies, primarily at $z\lesssim3$, have revealed a broad range of gas fractions \citep[e.g.,][]{Sargent2015,Suess2017,Spilker2018b,Gobat2018NatAs,Bezanson2019,Bezanson2022,Williams2021,Suzuki2022, Setton2025_SQuIGGL, Adscheid2025, Scholtz2026,  D_Eugenio2023, D_Eugenio2026}. 
While substantial molecular gas reservoirs have been reported in some quiescent systems \citep[e.g.,][]{Suess2017,Gobat2018NatAs}, others show stringent constraints well below those of coeval star-forming galaxies \citep[e.g.,][]{Spilker2018b,Bezanson2019,Williams2021, Scholtz2026}. This diversity leaves an open question of whether the depletion or removal of molecular gas is generally associated with the initial shutdown of star formation. Moreover, these systems are subject to substantial cosmic evolution and thus may not preserve the information of their quenching process cleanly. Most previous gas studies have targeted galaxies that are already quiescent at $z\lesssim3$, whose gas reservoirs may have continued to evolve after quenching through gas consumption, removal, heating, or subsequent accretion \citep[e.g.,][]{ Hartley2023,Lorenzon2025, Woodrum2022}. Their observed gas content may therefore not directly reflect the state of the gas reservoir at the time of quenching. Recently quenched dormant galaxies at $z>3$, in contrast, provide a more direct probe of the interstellar medium shortly after the onset of quenching. A substantial remaining molecular gas reservoir would favor a temporarily suppressed, low-SFE state, whereas an already depleted reservoir comparable to that of quiescent galaxies would suggest that the transition toward a gas-poor state can occur rapidly.

In this work, we combine JWST/NIRSpec spectroscopy with archival ALMA observations of \cii~158\,$\mu$m and CO rotational lines to constrain the molecular gas content of dormant galaxies\footnote{Throughout this work, ``dormant'' denotes galaxies selected using criteria similar to those of \citet{Covelo-Paz2026}, rather than implying a temporary cessation of star formation.} at $z>3$, alongside a comparison sample of quiescent galaxies. By systematically comparing their molecular gas fractions, we test whether dormant galaxies retain substantial molecular gas reservoirs shortly after quenching. Our primary goal is to distinguish whether their rapidly declining star formation is associated predominantly with the depletion/removal of molecular gas or with suppressed SFE, and ultimately to address whether these dormant galaxies are already becoming ``\textit{dead}'' or are temporarily ``\textit{sleeping}'', if no substantial gas replenishment occurs in the future.

This paper is organized as follows. In Sect.~\ref{Sec: data}, we describe the JWST parent sample, the selection of dormant and quiescent galaxies, and the ALMA observations and data reduction. In Sect.~\ref{Sec: gas}, we present the molecular gas measurements and derive molecular gas masses and gas fractions. Section~\ref{Sec: result} presents the evidence for recent quenching and the molecular gas fractions of dormant and quiescent galaxies. In Sect.~\ref{Sec: discussion}, we discuss the implications for gas depletion/removal versus suppressed star-formation efficiency and the subsequent fate of dormant galaxies. We summarize our conclusions in Sect.~\ref{Sec: conclusion}.

Throughout this paper, we adopt a Chabrier initial mass function \citep[IMF;][]{Chabrier2003} to estimate star formation rate (SFR) and stellar mass ($M_{\star}$). We assume a Planck cosmology \citep{Planck2020} with $(\Omega_\mathrm{m},\, \Omega_\mathrm{\Lambda},\, h)=(0.3,\, 0.7,\, 0.7)$. When necessary, data from the literature have been converted with a conversion factor of $M_{\star}$ \citep[][IMF]{Salpeter1955} = 1.7  $\times$ $M_{\star}$ \citep[][IMF]{Chabrier2003}. All magnitudes are in the AB system \citep{Oke1983}, such that $m_{\rm AB} = 23.9 - 2.5$ $\times$ log(S$_{\nu}$ [$\mu$Jy]).

\section{JWST sample selection and ALMA observations}\label{Sec: data}

\subsection{JWST parent sample}\label{Sect:parent}
\begin{figure*}[!t]
\centering
\includegraphics[width=18.5cm]{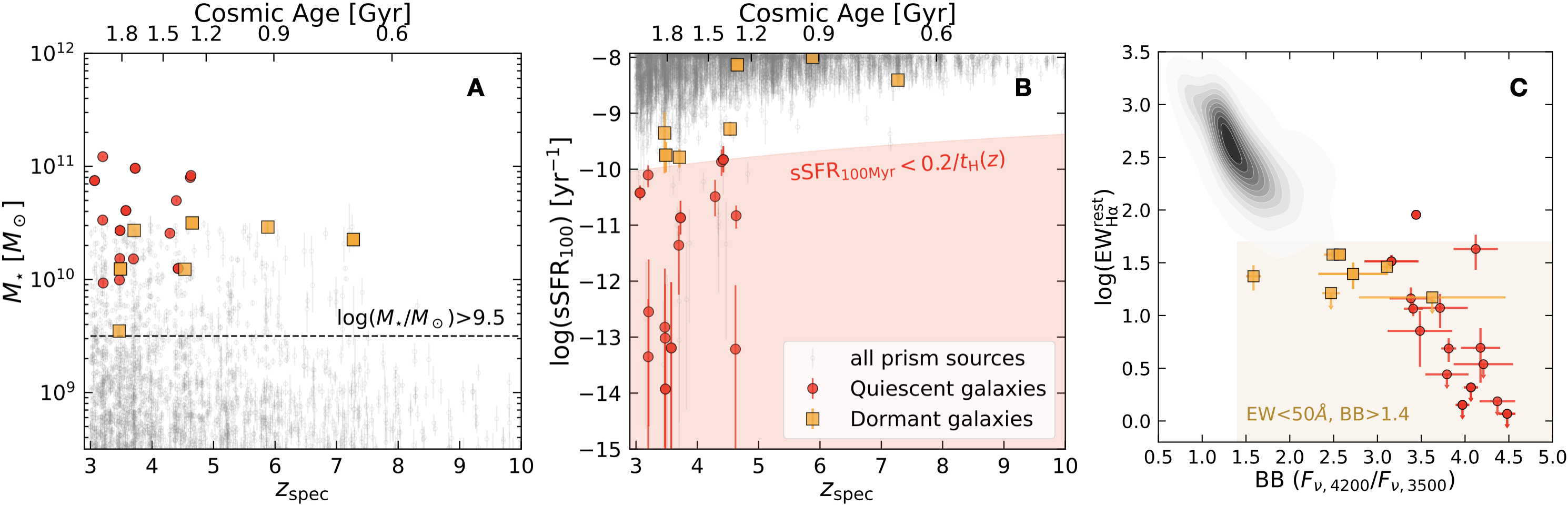}
\caption{
\textbf{Sample selection of dormant galaxies (orange) and quiescent galaxies (red).}
The full population of JWST/NIRSpec prism observations in the legacy fields that are accessible to ALMA (Dec $<20^\circ$) is shown in grey and serves as the parent sample. We first select galaxies with $\log(M_{\star}/M_\odot)>9.5$ at $z_{\rm spec}>3$ (panel~\textbf{A}). Quiescent galaxies (red circles) are selected using criterion~\#1, $\mathrm{sSFR}_{\rm 100\,Myr}<0.2/t_{\rm H}(z)$ (red shaded region in panel~\textbf{B}). Among the remaining galaxies that do not satisfy criterion~\#1, we identify "Dormant galaxies" (orange squares) using criterion~\#2: $F_{\nu,4200}/F_{\nu,3500}>1.4$, $A_V<1$\,mag, and $\mathrm{EW}_{\mathrm{H}\alpha}^{\rm rest}<50$\,\AA\ (yellow shaded region in panel~\textbf{C}). For more details, see Sect.~\ref{Sect:selection}. Error bars correspond to 1$\sigma$ uncertainties. Upper limits are 3$\sigma$.
}
\label{sample}
\end{figure*}

To build a JWST spectroscopic sample of dormant galaxies with ALMA constraints on their molecular gas content, we start from all publicly available JWST/NIRSpec prism spectra in the DAWN JWST Archive (DJA) v4.5 \citep{msaexp,deGraaff2025_rubies,Heintz2025, Valentino2025}\footnote{Latest DJA public spectra: \url{https://s3.amazonaws.com/msaexp-nirspec/extractions/nirspec_public_v4.5.html}}. We select sources with secure spectroscopic redshifts, $z_{\rm spec}>3$, located in the legacy fields that are accessible to ALMA (Dec $<20^\circ$), i.e., COSMOS, UDS, and GOODS-South\footnote{As a subsequent check, we also searched the lensed fields in the DJA v4.5 catalog for sources satisfying our QG and dormant galaxy selection criteria (Sect.~\ref{Sect:selection}), but found no additional sources with available ALMA observations.}. We further require stellar masses of $\log(M_{\star}/M_\odot)>9.5$ (Fig.~\ref{sample}-A). 
This mass cut is motivated by the limited depth of the archival ALMA observations: at lower stellar masses, even the molecular gas reservoirs expected for star-forming main sequence (SFMS) galaxies would generally be undetectable \citep[e.g.,][]{Liu2019,Dessauges-Zavadsky2020, tacconi2018, scoville2017}, leaving little constraint on the gas fraction ($M_{\rm mol}/M_{\star}$). We therefore focus on more massive systems, for which the existing ALMA data can provide meaningful constraints on their molecular gas content ($\sim70\%$ of our final sample; see Sect.~\ref{Sec: gas_fraction}).

The stellar population properties are adopted from the spectrophotometric SED fitting presented in \cite{gottumukkala2026}\footnote{The stellar population catalog of prism sources at $z>2$: \url{https://doi.org/10.5281/zenodo.22734795}} \citep[see also][]{Xiao2026_massivesample}. Briefly, the JWST/NIRSpec prism spectra, covering a wavelength range of $\sim0.5-5.5\,\mu$m, are fitted simultaneously with the available NIRCam broad-band photometry using \texttt{Bagpipes} \citep{Carnall2018,Carnall2019} at fixed $z_{\rm spec}$. The fitting adopts a non-parametric star-formation history (SFH) with a continuity prior, BPASS stellar population models, and the \citet{Salim2018} dust attenuation curve. The SFH is described by ten time bins. The first five bins are fixed at 0, 3, 10, 25, 50, and 100\,Myr, allowing us to separately trace star formation on recent 10\,Myr and 100\,Myr timescales (Figs.~\ref{ms},~\ref{sfr10_100}; Sect.~\ref{Sec: MS}). The remaining five bins are logarithmically spaced between 100\,Myr and the maximum allowed age, defined as the age of the Universe relative to its age at $z=30$.  The fitting returns posterior distributions for the $M_{\star}$, SFR, dust attenuation to the rest-frame V band ($A_V$), mass-weighted age (Age$_{\rm MW}$), and SFH, which are used throughout this work.

For a detailed assessment of the robustness of the derived stellar masses, including comparisons using different SFHs, dust attenuation curves, stellar population models, and fits with and without the prism spectra, we refer the reader to \citet{Xiao2026_massivesample}.

\subsection{Selection of quiescent galaxies and dormant galaxies}\label{Sect:selection}
We select quiescent galaxies (QGs) using an evolving specific star-formation rate criterion \citep[criterion \#1; Fig.~\ref{sample}-B; e.g.,][]{Carnall2024,Baker2026_twominiquench,Gallazzi2014},
\begin{equation}\label{eq:sSFR}
   \mathrm{sSFR}_{\rm 100\,Myr} < 0.2/t_{\rm H}(z),
\end{equation}
where $t_{\rm H}(z)$ is the cosmic age at the redshift of each galaxy. This criterion identifies galaxies with strongly suppressed star formation averaged over the past 100\,Myr. 

However, because $\mathrm{sSFR}_{\rm 100\,Myr}$ averages over a relatively long timescale, it can miss galaxies whose star formation has declined only very recently. We therefore define an additional class of ``dormant'' galaxies following a selection similar to that of \citet{Covelo-Paz2026}, using spectral indicators of a recent shutdown of star formation. Note that here, we use the term ``dormant'' as a label for galaxies satisfying this spectral selection, consistent with the terminology in the literature, rather than as an interpretation of their current or future star-formation activity.
Among galaxies that do not satisfy criterion \#1 (Fig.~\ref{sample}-B), we select dormant galaxies using criterion \#2 (Fig.~\ref{sample}-C):
\begin{align}\label{eq:dormant1}
F_{\nu,4200}/F_{\nu,3500} > 1.4 \quad \& \quad A_V < 1~{\rm mag}, \\
\mathrm{EW}_{\mathrm{H}\alpha}^{\rm rest} < 50\,{\rm \AA}.\label{eq:dormant2}
\end{align}
We calculate the Balmer break strength (BB; $F_{\nu,4200}/F_{\nu,3500}$) and the rest-frame \ha\ equivalent width ($\mathrm{EW}_{\mathrm{H}\alpha}^\mathrm{rest}$) directly by fitting the observed spectra (Sawarkar in prep.).   
Specifically, $F_{\nu,4200}/F_{\nu,3500}$ is calculated as the ratio of the $f_\nu$ values between two sides of the Balmer break, a blue window from $3400-3600\AA$ and a red window from $4150-4350\AA$. The $\mathrm{EW}_{\mathrm{H}\alpha}^\mathrm{rest}$ is calculated by dividing the \ha\, flux by the local continuum, and dividing by (1+$z$) to convert to the rest frame. Since \ha\ is blended with the \nii\ doublet at the low resolution of the NIRSpec PRISM, we apply a small correction and assume that 90\% of the blended flux stems from the \ha\, line. However, we caution that a few sources like \texttt{4233\_140707}, whose feature at $\sim$6500\,\AA\ is dominated by a blended \nii\ doublet rather than \ha\, from medium-resolution observations with the EXCELS survey \citep{Carnall2024}. In such cases, the true $\mathrm{EW}_{\mathrm{H}\alpha}^\mathrm{rest}$ would be lower than those reported in Table~\ref{table: stellar}. This could potentially cause us to miss some dormant galaxies, but would not affect whether the sources already in our sample satisfy the weak-\ha\ selection. 

It is noteworthy that we also explored whether a simple threshold of
$\mathrm{SFR}_{10\,\mathrm{Myr}}/\mathrm{SFR}_{100\,\mathrm{Myr}}<0.1$
could serve as an alternative criterion for selecting recently quenched galaxies. Among the parent sample with $\log(M_{\star}/M_\odot)>9.5$ at $z>3$, this simple threshold can successfully select eight dormant galaxies (satisfying Eq.~\ref{eq:dormant1} \& Eq.~\ref{eq:dormant2}; four of eight are observed by ALMA and are in our final sample; see Tables~\ref{alma},~\ref{table: stellar} and Fig.~\ref{sfr10_100}), but also six (dusty) star-forming galaxies (one of six is observed by ALMA). Inspection of these six galaxies shows that they generally exhibit either substantial dust attenuation ($A_V=1.4-2.5$\,mag) or strong H$\alpha$ emission ($\mathrm{EW}_{\mathrm{H}\alpha}^{\rm rest}=71-154$\,\AA), likely being contaminants. It indicates that a low $\mathrm{SFR}_{10\,\mathrm{Myr}}/\mathrm{SFR}_{100\,\mathrm{Myr}}$ ratio alone may not be sufficient to robustly identify recently quenched galaxies. We also found that using $\mathrm{SFR}_{3\,\mathrm{Myr}}/\mathrm{SFR}_{100\,\mathrm{Myr}}$ yields similar results. We therefore adopt the spectroscopic selection based on the $F_{\nu,4200}/F_{\nu,3500}$, $\mathrm{EW}_{\mathrm{H}\alpha}^\mathrm{rest}$, and $A_V$ throughout this work.

Hereafter, galaxies selected with criterion~\#1 are referred to as "Quiescent galaxies" and are shown as red circles (Figs.~\ref{sample} -- \ref{fgas}). Galaxies satisfying criterion~\#2 but not criterion~\#1 are referred to as "Dormant galaxies" and are shown as orange squares.

\begin{table*}
\caption{Summary of ALMA observations with emission lines of our sample.}   
\tiny      
\setlength{\tabcolsep}{4pt}
\centering
\renewcommand{\arraystretch}{1.2} 
\begin{threeparttable} 
 
\begin{tabular}{l l c c cc c cc}    
\hline\hline       

JWST ID & RA & DEC & $z_{\rm spec}$ & Line &  beam size & ALMA Project ID & $t_{\rm exp}$ [s]$^{*}$ & Other ID  \\
\hline  
\hline  
\multicolumn{6}{l}{Dormant galaxies}\\
\hline  
\texttt{1286\_171323} &53.139508 & -27.874153&3.4669$_{-0.0040}^{+0.0046}$& CO(4-3) & 1\farcs54 $\times$ 1\farcs20 & 2024.1.01213.S & 847 \\ 
\texttt{1286\_189781} & 53.087374& -27.840008&3.4877$_{-0.0023}^{+0.0029}$& CO(4-3) & 1\farcs08 $\times$ 0\farcs68 & 2021.1.00024.S &163  \\ 
 & &&& CO(6-5) & 1\farcs11 $\times$ 0\farcs84 & 2021.1.00024.S & 151.2  \\   
\texttt{2565\_18842} & 150.087256&2.396039& 3.7136$_{-0.0015}^{+0.0020}$& CO(4-3) & 0\farcs41 $\times$ 0\farcs27 & 2016.1.01001.S & 816\\   
\texttt{2565\_30519} & 34.503440&-5.173710&4.5354$_{-0.0012}^{+0.0016}$ & \cii & 0\farcs79 $\times$ 0\farcs76 & 2024.1.01744.S & 726\\  
\texttt{2198\_8777} & 53.108204&-27.825188 &4.6585$_{-0.0015}^{+0.0019}$& \cii & 0\farcs23 $\times$ 0\farcs20 & 2023.1.01016.S & 15967 & GS-9209$^{a}$ \\ 
&&&& CO(2-1) & 1\farcs15 $\times$ 0\farcs85 & 2024.1.01662.S & 13608 \\
\texttt{1214\_5403} &150.109947 &2.259070& 5.8650$_{-0.0036}^{+0.0035}$ & CO(6-5) & 1\farcs42 $\times$ 1\farcs29 & 2018.A.00037.S & 10191 \\
\texttt{6368\_5643}$^{\dagger}$ & 34.429616 & -5.112317 & 7.2728$_{-0.0023}^{+0.0028}$ & \cii & 1\farcs33 $\times$ 1\farcs16 & 2025.1.00092.S  & 5690 & RUBIES-UDS-QG-z7$^{b}$\\
\hline  
\multicolumn{6}{l}{Quiescent galaxies}\\
\hline 
\texttt{1180\_197911} &53.165314 &-27.814140& 3.0678$_{-0.0003}^{+0.0004}$& CO(3-2) & 0\farcs70 $\times$ 0\farcs48 & 2023.1.01296.S & 16178  & GS-10578$^{c}$\\ 
 & && & CO(5-4) & 0\farcs33 $\times$ 0\farcs30 & 2015.1.01205.S & 1754 \\  
 \texttt{2565\_12629} &34.255885 &-5.233871 &3.1997$_{-0.0016}^{+0.0018}$& CO(4-3)& 0\farcs65 $\times$ 0\farcs41 & 2023.1.01292.S & 7893 & $^{d}$\\
\texttt{2565\_13079} & 34.258906&-5.232347 &3.2032$_{-0.0007}^{+0.0009}$& CO(4-3) & 0\farcs65 $\times$ 0\farcs38 & 2023.1.01292.S  & 7893  & $^{d}$\\ 
\texttt{1215\_5935} & 34.254473 & -5.234889 &3.2091$_{-0.0032}^{+0.0029}$& CO(4-3) & 0\farcs65 $\times$ 0\farcs41 & 2023.1.01292.S &7893  \\
\texttt{2198\_6620} &53.078727 &-27.839608& 3.4730$_{-0.0031}^{+0.0025}$& CO(4-3) & 1\farcs54 $\times$ 1\farcs20 & 2024.1.01213.S & 847\\  
\texttt{1212\_518} &53.079715 &-27.838217 &3.4749$_{-0.0034}^{+0.0041}$& CO(4-3) & 0\farcs96 $\times$ 0\farcs67 & 2021.1.00024.S & 284  \\ 
\texttt{1286\_10026167} &53.082581 & -27.866803&3.4754$_{-0.0026}^{+0.0025}$& CO(4-3) & 1\farcs54 $\times$ 1\farcs20 & 2024.1.01213.S & 1693\\  
 & &&& CO(6-5) & 1\farcs03 $\times$ 0\farcs79 & 2024.1.01213.S & 1270 \\ 
\texttt{1212\_4682} & 53.113258&-27.869955&3.5853$_{-0.0021}^{+0.0022}$ & CO(4-3)& 1\farcs54 $\times$ 1\farcs22& 2024.1.01213.S & 968\\   
 & &&&CO(6-5)& 1\farcs01 $\times$ 0\farcs79 & 2024.1.01213.S & 726  \\  
\texttt{2565\_35168} & 34.485150 & -5.157807&3.6952$_{-0.0026}^{+0.0022}$& CO(4-3) & 0\farcs69 $\times$ 0\farcs64 & 2021.1.01291.S & 5111 \\  
\texttt{2565\_20115} & 150.061467&2.378686&3.7228$_{-0.0015}^{+0.0014}$ & \cii& 0\farcs54 $\times$ 0\farcs43& 2015.A.00026.S,2022.1.01604.S & 6895  & Jekyll$^{e}$ \\  
 & &&& CO(6-5) & 1\farcs51 $\times$ 1\farcs44 & 2021.1.00225.S & 35   \\  
\texttt{2565\_10559} & 150.071436&2.291179& 4.2946$_{-0.0026}^{+0.0028}$& CO(5-4)& 2\farcs29 $\times$ 1\farcs76& 2021.1.01005.S & 1058\\   
\texttt{b28} &34.280515 &-5.217214& 4.4016$_{-0.0025}^{+0.0031}$& \cii & 0\farcs96 $\times$ 0\farcs77 & 2024.1.01744.S & 1300  & $^{d}$\\  
\texttt{1212\_5360} & 53.062270 & -27.8750481&4.4408$_{-0.0013}^{+0.0017}$ & CO(4-3) & 1\farcs05 $\times$ 0\farcs83 & 2021.1.00024.S & 121   \\   
 & && & CO(5-4) & 1\farcs36 $\times$ 0\farcs98 & 2018.1.01841.S & 1845 \\ 
\texttt{4233\_140707} & 34.365084 & -5.148848 &4.6195$_{-0.0022}^{+0.0022}$ & \cii & 1\farcs33 $\times$ 0\farcs67 & 2013.1.01292.S & 73  & $^{d}$\\   
\texttt{4233\_149494} & 34.399676& -5.136348 &4.6281$_{-0.0014}^{+0.0015}$ & \cii & 1\farcs29 $\times$ 0\farcs68 & 2013.1.01292.S & 73   & $^{d}$\\   
\hline
\end{tabular}
\begin{tablenotes}
\item \textbf{Note:} $^{*}$ On-source integration time from the ALMA Science Archive. ${\dagger}$ The ALMA data is not public, the ALMA relevant values of beam size, \cii\, flux,  $L_{\rm \cii}$, $L^\prime_{\rm \cii}$, and $M_{\rm mol}$ (listed in Table~\ref{Mgas}) are directly taken from \cite{Valentino2026}. $^{a}$ \cite{Carnall2023Natur}; $^{b}$ \cite{Weibel2025}; $^{c}$ \cite{Scholtz2026}; $^{d}$ \cite{Carnall2024}; $^{e}$ \cite{Schreiber2018_JekyllHyde}. 
\end{tablenotes}
\label{alma} 
\end{threeparttable} 
\end{table*}

\begin{table*}
\caption{Stellar properties of our sample with JWST.}   
\tiny       
\centering
\renewcommand{\arraystretch}{1.2} 
\begin{threeparttable} 
 
\begin{tabular}{ll c ccc c cc cc c}    
\hline\hline       
JWST ID & log$(M_{\star})$  & log(sSFR$_{100}$) & log(sSFR$_{10}$) & SFR$_{10}$/SFR$_{100}$ & $A_{\rm V}$ & BB & EW(H$\alpha$) & $\beta_{\rm UV}$ & Age$_{\rm MW}$\\
&[log$(M_\odot)$] &[log(yr$^{-1}$)]   &[log(yr$^{-1}$)] & & [mag] & & [$\AA$] & & [Gyr]  \\
\hline  
\multicolumn{6}{l}{Dormant galaxies}\\
\hline  
  \texttt{1286\_171323}  & $9.54_{-0.04}^{+0.03}$   &  $-9.35_{-0.75}^{+0.39}$  & $-10.83_{-1.35}^{+0.51}$ & $0.03_{-0.04}^{+0.09}$ & $0.38_{-0.15}^{+0.13}$ & $3.63\pm0.84$ & $<45$ & $-1.39_{-0.11}^{+0.11}$ & $0.56_{-0.11}^{+0.14}$  \\
 \texttt{1286\_189781}  & $10.09_{-0.03}^{+0.03}$  &  $-9.75_{-0.30}^{+0.23}$  & $-9.89_{-0.16}^{+0.14}$  & $0.71_{-0.41}^{+0.56}$ & $0.54_{-0.07}^{+0.07}$ & $2.72\pm0.40$ & $25\pm7$ & $-0.75_{-0.11}^{+0.14}$ & $0.79_{-0.16}^{+0.13}$   \\
 \texttt{2565\_18842}   & $10.43_{-0.01}^{+0.01}$  &  $-9.78_{-0.19}^{+0.14}$  & $-9.82_{-0.09}^{+0.12}$  & $0.91_{-0.36}^{+0.45}$ & $0.47_{-0.02}^{+0.03}$ & $3.11\pm0.07$ & $29\pm4$ & $0.32_{-0.09}^{+0.10}$ & $0.27_{-0.02}^{+0.02}$  \\
 \texttt{2565\_30519}   & $10.09_{-0.01}^{+0.01}$  &  $-9.28_{-0.13}^{+0.13}$  & $-9.53_{-0.09}^{+0.09}$  & $0.56_{-0.18}^{+0.23}$ & $0.05_{-0.02}^{+0.02}$ & $2.49\pm0.10$ & $38\pm4$ & $-1.63_{-0.03}^{+0.03}$ & $0.22_{-0.01}^{+0.01}$ \\
 \texttt{2198\_8777}    & $10.50_{-0.01}^{+0.01}$  &  $-8.14_{-0.05}^{+0.04}$  & $-9.61_{-0.10}^{+0.11}$  & $0.03_{-0.01}^{+0.01}$ & $0.67_{-0.04}^{+0.04}$ & $2.57\pm0.03$ & $38\pm4$ & $-0.34_{-0.04}^{+0.04}$ & $0.13_{-0.01}^{+0.01}$ \\
 \texttt{1214\_5403}    & $10.46_{-0.05}^{+0.05}$  &  $-7.91_{-0.08}^{+0.07}$  & $-9.64_{-1.36}^{+0.49}$  & $0.02_{-0.02}^{+0.04}$ &  $1.13_{-0.10}^{+0.10}$$^*$ 
 & $1.59\pm0.09$ & $24\pm6$ & $-1.18_{-0.05}^{+0.05}$ & $0.03_{-0.00}^{+0.01}$ \\ 
\texttt{6368\_5643}$^{\dagger}$ & $10.35_{-0.04}^{+0.05}$  & $-8.40_{-0.12}^{+0.12}$ & $-10.13_{-1.19}^{+0.49}$ & $0.02_{-0.02}^{+0.04}$ & $0.81_{-0.12}^{+0.14}$ & $2.47\pm0.10$ & $16.30\pm16.35$ & $-0.60_{-0.07}^{+0.08}$ & $0.24_{-0.04}^{+0.03}$  \\
\hline  
\multicolumn{6}{l}{Quiescent galaxies}\\
\hline 
 \texttt{1180\_197911}  & $10.87_{-0.01}^{+0.00}$  &  $-10.43_{-0.12}^{+0.13}$ & $-9.84_{-0.04}^{+0.04}$  & $3.87_{-1.00}^{+1.41}$ & $0.49_{-0.01}^{+0.01}$ & $3.44\pm0.02$ & $90\pm4$ & $0.88_{-0.03}^{+0.03}$ & $0.41_{-0.01}^{+0.01}$  \\
  \texttt{2565\_12629}   & $11.09_{-0.01}^{+0.01}$  &  $-13.36_{-1.82}^{+1.03}$ & $-13.97_{-2.28}^{+1.42}$ & $0.24_{-0.34}^{+6.49}$ & $0.54_{-0.04}^{+0.03}$ & $4.37\pm0.20$ & $<5$ & $3.55_{-0.16}^{+0.12}$ & $1.39_{-0.00}^{+0.00}$   \\
 \texttt{2565\_13079}   & $10.53_{-0.02}^{+0.03}$  &  $-10.10_{-0.21}^{+0.19}$ & $-9.76_{-0.10}^{+0.11}$  & $2.19_{-0.95}^{+1.32}$ & $0.79_{-0.04}^{+0.04}$ & $4.12\pm0.25$ & $43\pm16$ & $1.26_{-0.31}^{+0.29}$ & $0.49_{-0.08}^{+0.14}$   \\
 \texttt{1215\_5935}    & $9.97_{-0.03}^{+0.02}$   &  $-12.55_{-2.08}^{+0.91}$ & $-12.87_{-1.99}^{+1.12}$ & $0.48_{-0.67}^{+6.74}$ & $0.15_{-0.06}^{+0.06}$ & $3.49\pm0.37$ & $7\pm4$ & $2.05_{-0.62}^{+0.46}$ & $1.36_{-0.05}^{+0.03}$   \\
  \texttt{2198\_6620}    & $10.19_{-0.02}^{+0.01}$  &  $-13.01_{-2.31}^{+0.98}$ & $-13.66_{-2.15}^{+1.21}$ & $0.23_{-0.32}^{+3.97}$ & $0.13_{-0.03}^{+0.02}$ & $4.21\pm0.34$ & $<10$ & $2.35_{-0.33}^{+0.25}$ & $1.28_{-0.02}^{+0.02}$  \\
 \texttt{1212\_518}     & $10.00_{-0.01}^{+0.01}$  &  $-12.83_{-2.14}^{+1.06}$ & $-13.20_{-2.24}^{+1.22}$ & $0.43_{-0.60}^{+8.11}$ & $0.01_{-0.01}^{+0.01}$ & $3.71\pm0.32$ & $12\pm4$ & $0.06_{-0.06}^{+0.04}$ & $1.07_{-0.02}^{+0.02}$   \\
 \texttt{1286\_10026167}& $10.43_{-0.01}^{+0.01}$  &  $-13.93_{-2.18}^{+1.14}$ & $-14.34_{-2.10}^{+1.29}$ & $0.38_{-0.54}^{+8.58}$ & $0.00_{-0.00}^{+0.01}$ & $3.97\pm0.08$ & $<4$ & $0.43_{-0.04}^{+0.04}$ & $1.19_{-0.01}^{+0.01}$  \\
 \texttt{1212\_4682}    & $10.61_{-0.01}^{+0.01}$  &  $-13.19_{-2.04}^{+1.19}$ & $-13.45_{-2.05}^{+1.08}$ & $0.55_{-0.77}^{+10.05}$& $0.18_{-0.02}^{+0.02}$ & $4.07\pm0.08$ & $<6$ & $1.11_{-0.07}^{+0.08}$ & $0.41_{-0.00}^{+0.01}$ \\
 \texttt{2565\_35168}   & $10.18_{-0.02}^{+0.02}$  &  $-11.36_{-0.91}^{+0.37}$ & $-11.21_{-1.40}^{+0.36}$ & $1.42_{-1.85}^{+2.65}$ & $0.01_{-0.01}^{+0.02}$ & $3.38\pm0.24$ & $15\pm4$ & $-0.31_{-0.17}^{+0.18}$ & $0.91_{-0.04}^{+0.04}$   \\
 \texttt{2565\_20115}   & $10.98_{-0.01}^{+0.01}$  &  $-10.86_{-0.30}^{+0.30}$ & $-10.65_{-0.28}^{+0.17}$ & $1.63_{-1.12}^{+1.78}$ & $0.51_{-0.02}^{+0.02}$ & $4.48\pm0.09$ & $<4$ & $2.48_{-0.22}^{+0.21}$ & $0.41_{-0.00}^{+0.00}$  \\
 \texttt{2565\_10559}   & $10.41_{-0.02}^{+0.02}$  &  $-10.48_{-0.37}^{+0.30}$ & $-10.49_{-0.25}^{+0.21}$ & $0.98_{-0.71}^{+1.14}$ & $0.54_{-0.05}^{+0.05}$ & $3.79\pm0.25$ & $<8$ & $0.31_{-0.14}^{+0.15}$ & $0.37_{-0.02}^{+0.02}$  \\
 \texttt{b28}           & $10.70_{-0.01}^{+0.02}$  &  $-9.87_{-0.25}^{+0.23}$  & $-10.00_{-0.20}^{+0.18}$ & $0.74_{-0.42}^{+0.63}$ & $0.79_{-0.04}^{+0.04}$ & $4.18\pm0.22$ & $5\pm3$ & $1.81_{-0.29}^{+0.28}$ & $0.23_{-0.00}^{+0.01}$ \\
 \texttt{1212\_5360}    & $10.10_{-0.01}^{+0.02}$  &  $-9.83_{-0.23}^{+0.24}$  & $-9.69_{-0.10}^{+0.11}$  & $1.37_{-0.62}^{+1.07}$ & $0.61_{-0.03}^{+0.04}$ & $3.16\pm0.31$ & $33\pm5$ & $0.14_{-0.11}^{+0.12}$ & $0.39_{-0.02}^{+0.02}$\\
 \texttt{4233\_140707}  & $10.90_{-0.01}^{+0.01}$  &  $-13.20_{-1.89}^{+1.13}$ & $-13.38_{-2.16}^{+1.42}$ & $0.67_{-0.94}^{+18.85}$& $0.29_{-0.02}^{+0.02}$ & $3.41\pm0.11$ & $12\pm2$ & $3.07_{-0.29}^{+0.13}$ & $1.04_{-0.00}^{+0.00}$\\
 \texttt{4233\_149494}  & $10.92_{-0.01}^{+0.01}$  &  $-10.83_{-0.23}^{+0.19}$ & $-10.52_{-0.12}^{+0.12}$ & $2.05_{-0.96}^{+1.30}$ & $0.52_{-0.03}^{+0.03}$ & $3.81\pm0.09$ & $5\pm1$ & $1.46_{-0.18}^{+0.21}$ & $0.70_{-0.02}^{+0.02}$ \\
\hline
\end{tabular}
\begin{tablenotes}
\item \textbf{Note:} Measurements of physical properties from \cite{gottumukkala2026} \citep[see also][]{Xiao2026_massivesample}. BB and EW(H$\alpha$) are taken from Sawarkar in prep. We use 1$\sigma$ for uncertainties and 3$\sigma$ upper limits for non-detections. 
$^*$ This source was selected in a previous fitting run with $A_V=1.06_{-0.08}^{+0.07}$ mag, consistent within $1\sigma$ uncertainty with both the current estimate listed here and the dormant galaxy criterion of $A_V<1$ mag (Eq.~\ref{eq:dormant1}).
We therefore retain it in our sample.
\end{tablenotes}
\label{table: stellar} 
\end{threeparttable} 
\end{table*}

\subsection{ALMA observations and data reduction}\label{Sect:alma}
Applying criteria~\#1 and \#2 yields 54 quiescent or dormant galaxies. We then systematically searched the ALMA Science Archive\footnote{ALMA Science Archive: \url{https://almascience.eso.org/aq/}} for observations covering the expected frequencies of \cii~158\,$\mu$m or CO rotational transitions ($1\leq J_{\rm up}\leq6$) at the JWST spectroscopic redshifts of these galaxies. We considered an ALMA observation to be matched if the source lies within one primary beam size of the ALMA position. We focus on CO transitions up to $J_{\rm up}=6$, since higher-$J$ transitions preferentially trace warmer and denser gas and are therefore less direct tracers of the total molecular gas reservoir.

Among the 54 galaxies, 22 have archival ALMA observations covering either \cii~158\,$\mu$m or a CO rotational transition.  
This sample includes RUBIES-UDS-QG-z7, whose ALMA data are not yet public; for this galaxy, we adopt the published \cii\ measurement from \citet{Valentino2026}. The ALMA observations used in this work are summarized in Table~\ref{alma}, while the corresponding JWST-derived stellar properties are listed in Table~\ref{table: stellar}.

Data calibration of each program was performed using the standard ALMA pipeline within the Common Astronomy Software Application package \citep[CASA;][]{McMullin2007}.  We then performed further analysis with version 6.7.0.31 of CASA. When a source has multiple observations of the same transition with comparable spatial and spectral setups, we combine the calibrated measurement sets using \texttt{concat}. In cases where the observations have substantially different angular resolutions or spectral setups, we inspect the datasets separately and use the deepest observation for the final flux constraint if no significant detection is found.

Continuum subtraction is performed in the visibility plane with \texttt{uvcontsub}, where a first-order polynomial is fitted to the line-free channels identified by excluding a velocity window around the expected line center, determined from the JWST spectroscopic redshift. We then image the continuum-subtracted data cubes using \texttt{tclean}, adopting natural weighting to maximize point-source sensitivity. The pixel size is set to be 0.2 times the synthesized beam size (i.e., the angular resolution of the observation), and the channel width is set to 30-50\,km\,s$^{-1}$. The resulting data cubes are used for the line-flux measurements and upper-limit estimates described in Sect.~\ref{Sec: gas}.

\begin{table*}
\caption{Gas properties of our sample with ALMA.}
\tiny          
\centering
\renewcommand{\arraystretch}{1.2} 
\begin{threeparttable} 
 
\begin{tabular}{l c ccc c cc cc c}    
\hline\hline       
 JWST ID & Line & Flux$^{a}$ & $L_{\rm line}$ & $L^\prime_{\rm line}$ & $L^\prime_{\rm CO(1-0)}$ & $M_{\rm mol}$$^{b}$ & $\mu_{\rm mol}$$^{c}$ \\ 
&& [Jy km s$^{-1}$] & [$10^{8}\,L_\odot$] & [$10^{10}$ K km s$^{-1}$ pc$^2$] & [$10^{10}$ K km s$^{-1}$ pc$^2$] & [$10^{10}\,M_\odot$]  \\  
\hline  
\multicolumn{6}{l}{Dormant galaxies}\\
\hline  
 \texttt{1286\_171323}  & CO(4-3)          & $<$0.203  &   $<$0.200 & $<$0.64  & $<$1.04 & $<$4.55 & $<$13.0  \\
 \texttt{1286\_189781}  & CO(4-3)          & $<$0.271  &   $<$0.269 & $<$0.86  & $<$1.40 & $<$6.12 & $<$4.94  \\
                        & CO(6-5)          & $<$0.291  &   $<$0.432 & $<$0.41  & $<$1.46 & $<$6.37 & $<$5.14  \\
 \texttt{2565\_18842}   & CO(4-3)          & $<$0.136  &   $<$0.149 & $<$0.48  & $<$0.78 & $<$3.40 & $<$1.25  \\
 \texttt{2565\_30519}   & \cii             & $<$0.186  &   $<$1.149 & $<$0.05  & ...     & $<$0.36 & $<$0.29\\
 \texttt{2198\_8777}    & \cii             & $<$0.043  &   $<$0.276 & $<$0.01  & ...     & $<$0.09 & $<$0.03\\
                        & CO(2-1)          & $<$0.033  &   $<$0.026 & $<$0.66  & $<$0.68 & $<$2.95 & $<$0.94\\
 \texttt{1214\_5403}    & CO(6-5)          & $<$0.030  &   $<$0.098 & $<$0.09  & $<$0.33 & $<$1.44 & $<$0.50\\ 
 \texttt{6368\_5643}$^{\dagger}$    & \cii & $0.08_{-0.02}^{+0.02}$ & $1.1_{-0.2}^{+0.3}$ & $0.05_{-0.01}^{+0.01}$ & ... & $0.34_{-0.17}^{+0.37}$ & $0.15_{-0.08}^{+0.17}$  \\
\hline  
\multicolumn{6}{l}{Quiescent galaxies}\\
\hline 
 \texttt{1180\_197911}  & CO(3-2)          & $<$0.033  &   $<$0.020 & $<$0.15  & $<$0.19 & $<$0.82 & $<$0.11 \\
                        & CO(5-4)          & $<$0.122  &   $<$0.123 & $<$0.20 & $<$0.45 & $<$1.98 & $<$0.26  \\
 \texttt{2565\_12629}   & CO(4-3)          & $<$0.050  &   $<$0.043 & $<$0.14  & $<$0.23 & $<$0.99 & $<$0.08  \\
 \texttt{2565\_13079}   & CO(4-3)          & $<$0.049  &   $<$0.043 & $<$0.14  & $<$0.22 & $<$0.97 & $<$0.29  \\
 \texttt{1215\_5935}    & CO(4-3)          & $<$0.058  &   $<$0.050 & $<$0.16  & $<$0.26 & $<$1.15 & $<$1.23  \\
 \texttt{2198\_6620}    & CO(4-3)          & $<$0.264  &   $<$0.260 & $<$0.83  & $<$1.36 & $<$5.92 & $<$3.86  \\
 \texttt{1212\_518}     & CO(4-3)          & $<$0.259  &   $<$0.255 & $<$0.81  & $<$1.33 & $<$5.79 & $<$5.86  \\
 \texttt{1286\_10026167}& CO(4-3)          & $<$0.120  &   $<$0.118 & $<$0.38  & $<$0.62 & $<$2.69 & $<$0.99  \\
                        & CO(6-5)          & $<$0.096  &   $<$0.141 & $<$0.13  & $<$0.48 & $<$2.08 & $<$0.77  \\
 \texttt{1212\_4682}    & CO(4-3)          & $<$0.163  &   $<$0.169 & $<$0.54  & $<$0.88 & $<$3.85 & $<$0.94 \\
                        & CO(6-5)          & $<$0.236  &   $<$0.366 & $<$0.35  & $<$1.24 & $<$5.39 & $<$1.32  \\
 \texttt{2565\_35168}   & CO(4-3)          & $<$0.078  &   $<$0.084 & $<$0.27  & $<$0.44 & $<$1.92 & $<$1.26  \\
 \texttt{2565\_20115}   & \cii             & $<$0.130  &   $<$0.590 & $<$0.03  & ...     & $<$0.18 & $<$0.02 \\
                        & CO(6-5)          & $<$0.660  &   $<$1.088 & $<$1.03  & $<$3.67 & $<$16.02& $<$1.66 \\
 \texttt{2565\_10559}   & CO(5-4)          & $<$0.133  &   $<$0.230 & $<$0.37  & $<$0.85 & $<$3.71 & $<$1.45 \\
 \texttt{b28}           & \cii             & $<$0.313  &   $<$1.481 & $<$0.07  & ...     & $<$0.46 & $<$0.09\\
 \texttt{1212\_5360}    & CO(4-3)          & $<$0.381  &   $<$0.552 & $<$1.76  & $<$2.89 & $<$12.59& $<$10.07\\
                        & CO(5-4)          & $<$0.166  &   $<$0.301 & $<$0.49  & $<$1.12 & $<$4.86 & $<$3.89\\
 \texttt{4233\_140707}  & \cii             & $<$1.090  &   $<$6.927 & $<$0.32  & ...     & $<$2.15 & $<$0.27\\
 \texttt{4233\_149494}  & \cii             & $<$3.310  &   $<$21.095 & $<$0.96 & ...     & $<$6.54 & $<$0.79\\
\hline
\end{tabular}
\begin{tablenotes}
\item \textbf{Note:} We use 1$\sigma$ for uncertainties and 3$\sigma$ upper limits for non-detections. (a) Sources are not detected; we adopt 3$\sigma$ upper limits assuming line width of 250\,km\,s$^{-1}$ (see Sect.~\ref{Sec: gas}).    
(b) $M_{\rm mol}$ is calculated assuming $\alpha_{\rm \cii}=31\, M_{\odot}/L_{\odot}$ for \cii\, and $\alpha_{\rm CO}=4.36~M_{\odot}\,({\rm K\,km\,s^{-1}\,pc^{2}})^{-1}$ for CO (see Sect.~\ref{Sec: gas}). (c) Molecular gas fraction: $\mu_{\rm mol}={M_{\rm mol}}/{M_{\star}}$. ${\dagger}$ The ALMA data is not public, the \cii\, flux,  $L_{\rm \cii}$, $L^\prime_{\rm \cii}$, and $M_{\rm mol}$ are directly taken from \cite{Valentino2026}.
\end{tablenotes}
\label{Mgas} 
\end{threeparttable} 
\end{table*}

\section{Molecular gas measurements}\label{Sec: gas}

We searched for both \cii~158\,$\mu$m and CO rotational transitions ($1\leq J_{\rm up} \leq6$) in the continuum-subtracted ALMA data cubes. Spectra were extracted at the JWST positions of the galaxies using apertures of the synthesized beam size, assuming that the emission is unresolved and centered at the JWST spectroscopic redshift. We found none of the archival datasets yield a significant ($>3\sigma$) line detection for our sample. Therefore, we derive 3$\sigma$ upper limits on the integrated line fluxes. The only exception is RUBIES-UDS-QG-z7, for which the ALMA data are not yet public, we directly adopt the published \cii\ line flux from \citet{Valentino2026}.

For the non-detections, we assume a fiducial full width at half maximum (FWHM) of the emission line of 250\,km\,s$^{-1}$, representative of massive galaxies at these redshifts \citep[e.g.,][]{Bethermin2020,Endsley2022}. Moment-0 maps are constructed by integrating over channels within $1.2\times\text{FWHM}$ (i.e. $\pm$ 150 km s$^{-1}$, enclosing 84\% of the flux for a Gaussian line profile). The flux uncertainty is estimated by applying a total-flux correction factor of 1/0.84 to the standard deviation of fluxes in 1000 beam-sized apertures randomly placed on the sigma-clipped moment-0 maps. We refer to Qi et al. in prep. for more technical details. The resulting 3$\sigma$ upper limits on the integrated line fluxes are listed in Table~\ref{Mgas}.

The integrated line fluxes are converted into line luminosities following \citet{Solomon1997}:
\begin{align}
    \label{LCII}
    & L_{\text{line}} = 1.04\cdot10^{-3} S_{\text{line}} \Delta V \nu_{\text{rest}} (1+z)^{-1} D_L^2 \,, \\
    & L'_{\text{line}} = 3.25\cdot10^7 S_{\text{line}} \Delta V \nu_{\text{obs}}^{-2} D_L^2 (1+z)^{-3} \,,
\end{align}
where $S\Delta V$ is the velocity-integrated line flux in Jy\,km\,s$^{-1}$, $\nu_{\rm rest}$ and $\nu_{\rm obs}$ are the rest-frame and observed frequencies in GHz, respectively, and $D_{\rm L}$ is the luminosity distance in Mpc. The luminosity $L_{\text{line}}$ is expressed in $L_\odot$, while the luminosity $L'_{\text{line}}$ expressed in K\,km\,s$^{-1}$\,pc$^{2}$.

We estimate the molecular gas mass ($M_{\text{mol}}$) using different calibrations for \cii\ and CO observations. For \cii, we adopt the empirical calibration of \citet{zanella2018} \citep[verified also in][]{Dessauges-Zavadsky2020, Dessauges-Zavadsky2026},
\begin{equation}
M_{\rm mol}=\alpha_{\rm \cii}\,L_{\rm \cii},
\end{equation}
where the \cii-to-H$_2$ conversion factor, $\alpha_{\rm \cii}=31\, M_{\odot}/L_{\odot}$, was empirically calibrated using star-forming galaxies over the broad redshift range $0<z<6$ and has an intrinsic scatter of 0.3 dex. We adopt this calibration to maintain consistency with recent high-redshift studies of (recently) quenched galaxies \citep[e.g.,][]{Valentino2026,D_Eugenio2026}, enabling direct comparisons in the following section. Note that the ionization potential of carbon is 11.26\,eV, meaning that the \cii\ line can arise from ionized, atomic, and molecular gas. Throughout this work, we use the above empirical calibration, focusing only on the molecular gas content.

For CO observations, we first convert the observed transition to the equivalent CO(1-0) luminosity using the average excitation ladder $r_{J,1}$ and corresponding uncertainties from \citet{boogaard2020}. 
The molecular gas mass is then calculated assuming a Milky Way CO-to-H$_2$ conversion factor of $\alpha_{\rm CO}=4.36~M_{\odot}\,({\rm K\,km\,s^{-1}\,pc^{2}})^{-1}$ \citep[e.g.,][]{Tacconi2020}: 
\begin{align}
    \label{Mmolgas}
    & M_{\rm mol} = \alpha_{\text{CO}}\frac{
      L'_{\text{CO(J,J-1)}}}{r_{J,1}} \,.
\end{align}
This has included a factor of 1.36 to account for heavy elements, primarily helium \citep[see review in][]{Bolatto2013_review}. Our adoption of a Milky Way $\alpha_{\rm CO}$ is motivated by the relatively high stellar masses of our sample: most galaxies have $\log(M_{\star}/M_\odot)\gtrsim10$ (median of $\log(M_{\star}/M_\odot)=10.4$). At this typical mass, recent observational and simulated mass-metallicity relations suggest gas-phase metallicities of $12+\log(\mathrm{O/H})\gtrsim 8.4$ at $z\sim3-5$ \citep[e.g.,][]{Sharda2026, Faisst2026, Khostovan2026}, broadly consistent with a Milky Way $\alpha_{\rm CO}$ given the $\alpha_{\rm CO}$-metallicity relation \citep{maiolino2019}. 
The remaining uncertainty associated with $\alpha_{\rm CO}$ is discussed in Sect.~\ref{Sec: Caveats}.

Finally, we calculate the molecular gas fraction as
\begin{equation}
\mu_{\rm mol}=\frac{M_{\rm mol}}{M_{\star}},
\end{equation}
using the stellar masses derived from the joint JWST SED fitting (Sect.~\ref{Sect:parent}). All uncertainties arising from the line luminosities, [C \textsc{ii}]-to-H$_2$ \citep[0.3 dex;][]{zanella2018}  and CO-to-H$_2$ conversion factors \citep[30\%;][]{Bolatto2013_review}, the CO ladder \citep{boogaard2020}, and stellar mass were propagated to obtain the final uncertainty on $M_{\text{mol}}$ and $\mu_{\rm mol}$. The resulting molecular gas masses and gas fractions are summarized in Table~\ref{Mgas}.

\begin{figure*}
\centering
\includegraphics[width=14cm]{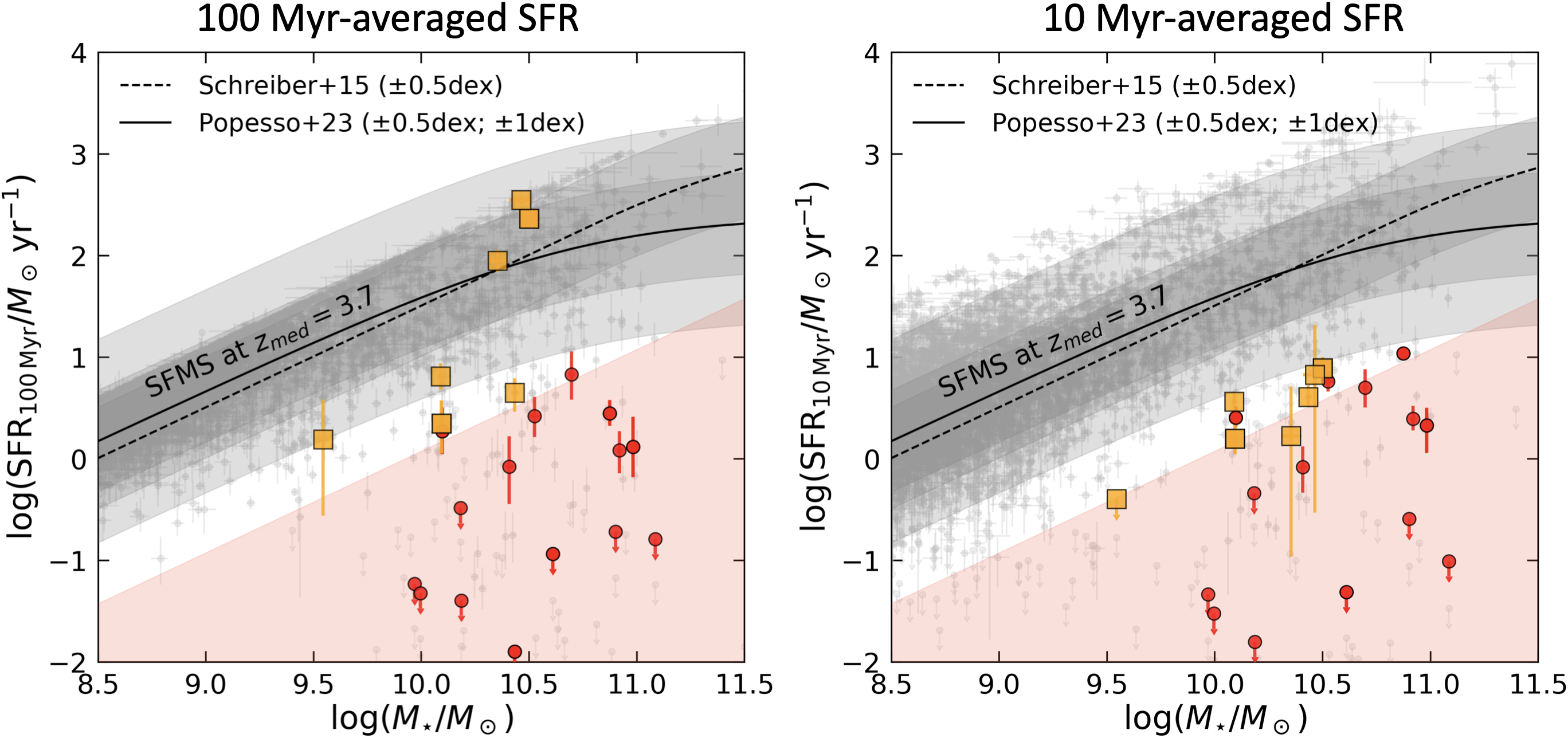}
\caption{
\textbf{Location of dormant galaxies (orange) and quiescent (red), together with all prism sources (grey), in the SFR-$M_{\star}$ plane.}
The full population of JWST/NIRSpec prism observations at $z_{\rm spec}>3$ in the legacy fields that are accessible to ALMA (Dec $<20^\circ$) is shown in grey points and serves as the parent sample. 
The SFMS relations of \citet{Popesso2023} and \citet{Schreiber2015} at $z=3.7$ (median redshift of our sample) are shown as black solid and dashed lines, respectively, with dark grey shading indicating their scatter of $\pm0.5$ dex. Light grey shading shows a more extended scatter of $\pm1$ dex around the \citet{Popesso2023} relation.
\textbf{Left}: Stellar mass versus SFR averaged over the past 100\,Myr. Most dormant galaxies remain close to the SFMS, although several already fall more than $\sim$1 dex below it, while quiescent galaxies lie well below the SFMS and satisfy the evolving $\mathrm{sSFR}_{\rm 100\,Myr}$ selection criterion \citep[red shaded region, shown at a representative median redshift of $z=3.7$; e.g.,][]{Valentino2023}.
\textbf{Right}: Stellar mass versus SFR averaged over the past 10\,Myr. Compared to the 100\,Myr-averaged SFRs, dormant galaxies shift systematically toward lower SFRs, with several approaching the quiescent galaxy population.
}
\label{ms}
\end{figure*}

\section{Results}\label{Sec: result}
\subsection{Evidence for recent quenching in dormant galaxies}\label{Sec: MS}

As described in Sect.~\ref{Sect:selection}, we refer to these galaxies as ``dormant'' because they are selected using spectral criteria similar to those adopted by \citet{Covelo-Paz2026}. Here, we investigate the recent star-formation activity of this selected population using their inferred SFR. Figure~\ref{ms} compares the locations of dormant galaxies and quiescent galaxies in the SFR-$M_{\star}$ plane using SFRs averaged over two different timescales. When the SFR is averaged over the past 100\,Myr (left panel), most dormant galaxies lie close to the SFMS, although two of them already fall more than $\sim$1 dex below it. In contrast, quiescent galaxies occupy the region well ($\gtrsim$1.5 dex) below the SFMS by construction (Eq.~\ref{eq:sSFR}). A different picture emerges when considering the SFR averaged over the past 10\,Myr (right panel). Relative to their positions in the left panel, dormant galaxies shift systematically toward lower SFRs, while the locations of quiescent galaxies remain largely unchanged. This indicates that dormant galaxies generally have experienced a rapid decline in star formation within the past $\sim$10\,Myr, despite having formed stars at substantially higher rates over the preceding $\sim$100\,Myr.

\begin{figure}
\centering
\includegraphics[width=9cm]{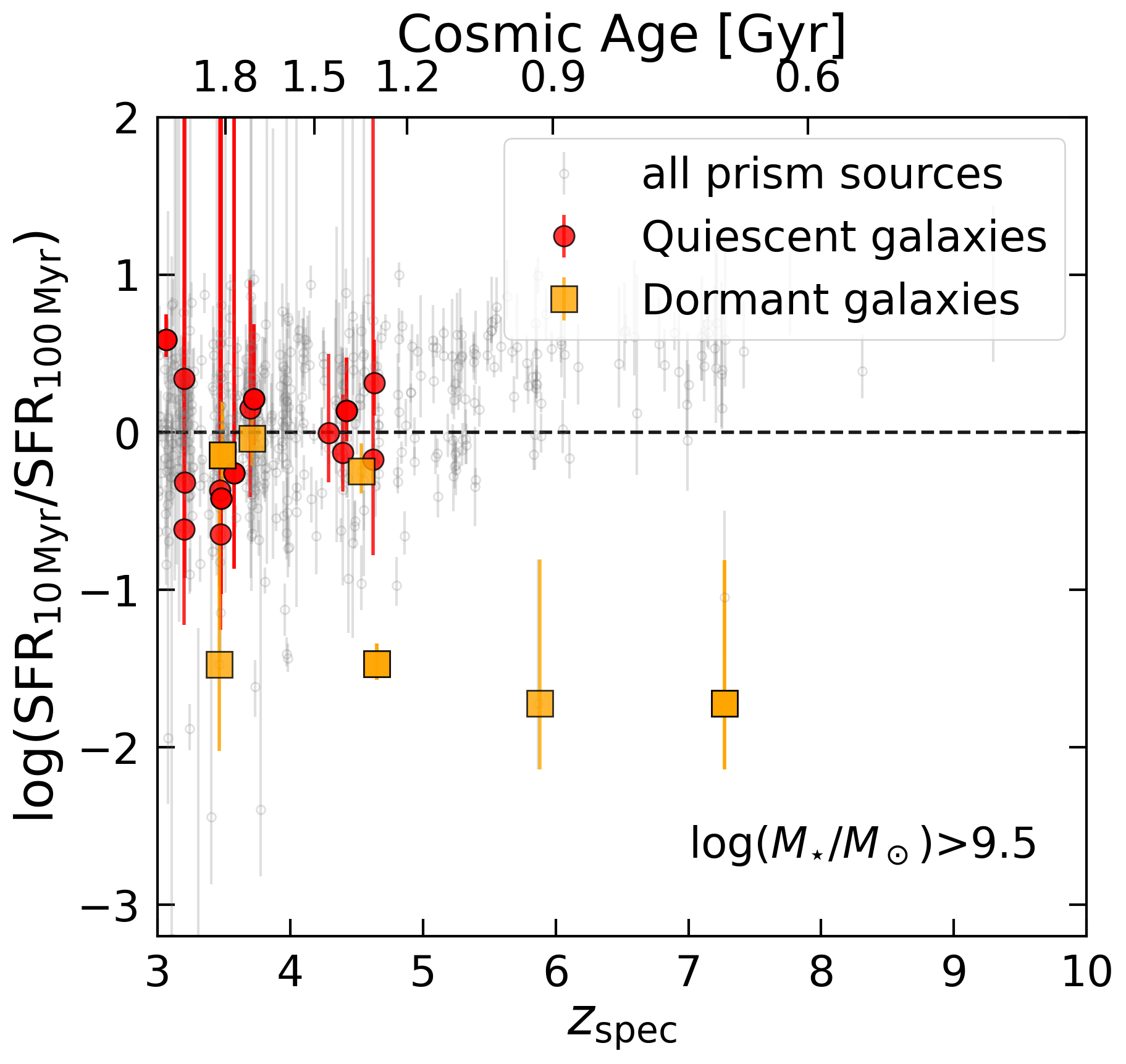}
\caption{
\textbf{SFR$_{\rm 10\,Myr}$/SFR$_{\rm 100\,Myr}$ ratios for dormant galaxies (orange) and quiescent galaxies (red) as a function of $z_{\rm spec}$.}
All JWST/NIRSpec prism sources with $\log(M_{\star}/M_\odot)>9.5$ at $z_{\rm spec}>3$ in the COSMOS, UDS, and GOODS-South fields are shown in grey for comparison. Dormant galaxies generally exhibit low $\mathrm{SFR}_{\rm 10\,Myr}/\mathrm{SFR}_{\rm 100\,Myr}$ ratios relative to the parent population, indicating a recent decline in star formation. For quiescent galaxies, both SFR estimates are low and have large uncertainties, leaving their ratios poorly constrained. The dashed horizontal line marks $\mathrm{SFR}_{\rm 10\,Myr}=\mathrm{SFR}_{\rm 100\,Myr}$.
}
\label{sfr10_100}
\end{figure}

The same trend is illustrated in Fig.~\ref{sfr10_100}, which shows the ratio between the SFR averaged over the past 10\,Myr and that averaged over the past 100\,Myr. Compared to both the parent population and quiescent galaxies, dormant galaxies, in general, preferentially occupy lower $\mathrm{SFR}_{10\,\mathrm{Myr}}/\mathrm{SFR}_{100\,\mathrm{Myr}}$ ratios, demonstrating that their recent star formation has declined significantly relative to their longer-term average.

Together, these results indicate that dormant galaxies selected in our sample generally represent systems caught shortly after the onset of quenching. This is also reflected in their individual SFH (see Appendix~\ref{appendix_sed}).
Their rapidly declining star formation activity makes them an ideal population for investigating the physical mechanisms responsible for shutting down star formation, which we explore through their molecular gas content in the following section.

\begin{figure*}
\centering
\includegraphics[width=18cm]{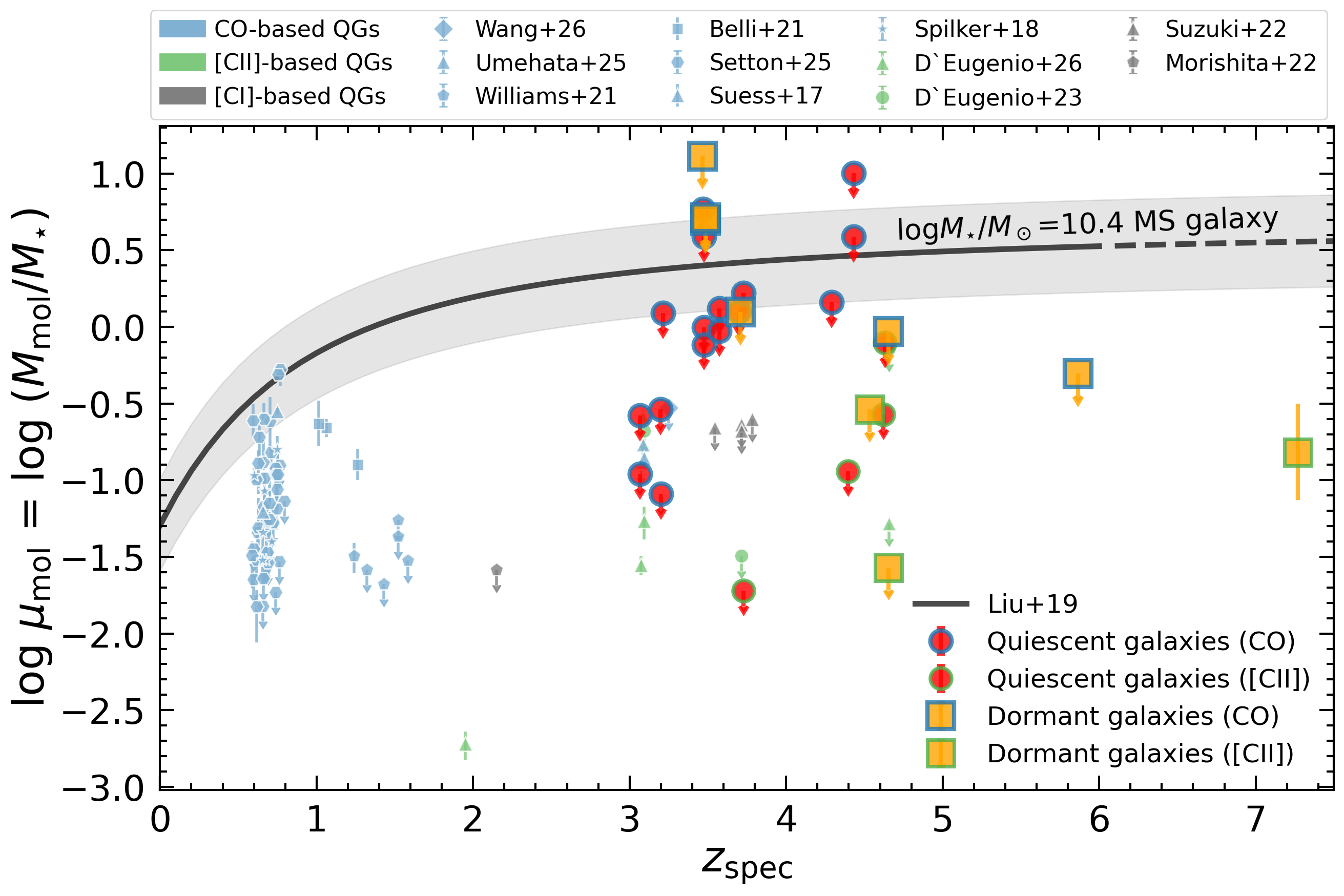}
\caption{
\textbf{Molecular gas fraction as a function of redshift}. Quiescent and dormant galaxies in our sample are shown as red circles and orange squares, respectively, with blue and green edges indicating estimates based on CO and \cii. The black solid line and shaded region show the \citet{Liu2019} scaling relation for main-sequence galaxies at the median stellar mass of our sample ($\log(M_{\star}/M_\odot)=10.4$), with a 0.3 dex scatter. Literature quiescent galaxies are shown with different symbols, color-coded by gas tracer (CO: blue; \cii: green; \ci: grey), as indicated in the legend. These measurements \citep{Suess2017, Spilker2018b, Williams2021, Belli2021, Setton2025_SQuIGGL, Umehata2025, Wang2026_redpotato, D_Eugenio2023, D_Eugenio2026, Suzuki2022, Morishita2022} are corrected, where necessary, to adopt the same conversion factors as our sample: $\alpha_{\rm \cii}=31\,M_{\odot}/L_{\odot}$ and $\alpha_{\rm CO}=4.36\,M_{\odot}\,({\rm K\,km\,s^{-1}\,pc^{2}})^{-1}$. Upper limits are 3$\sigma$. 
}
\label{fgas}
\end{figure*}

\subsection{Molecular gas fractions}\label{Sec: gas_fraction}
Figure~\ref{fgas} shows the molecular gas fractions of dormant galaxies and quiescent galaxies as a function of spectroscopic redshift. For our sample, all measurements are 3$\sigma$ upper limits, except for RUBIES-UDS-QG-z7, for which we adopt the published \cii-based molecular gas measurement and 1$\sigma$ uncertainty from \citet{Valentino2026}.

The depths of the available ALMA archival observations vary substantially from source to source, resulting in molecular gas fraction upper limits spanning nearly three orders of magnitude. Approximately 30\% of the sample have upper limits that lie on or above the expected molecular gas fractions of main-sequence galaxies \citep[$\pm$0.3 dex; at the median stellar mass of our sample, i.e., $\log(M_{\star}/M_\odot)=10.4$;][]{Liu2019}, owing to the relatively shallow ALMA observations. 
Therefore, their molecular gas reservoirs remain poorly constrained with the current ALMA data. Nevertheless, for most of our sample ($\sim70\%$), the current observations can place meaningful constraints on their molecular gas fractions, with upper limits $\gtrsim$0.3 dex below the typical values expected for main-sequence galaxies.

Figure~\ref{fgas} reveals two main results. Among the $\sim70\%$ of the sample with sufficiently constraining ALMA observations: First, the detections or 3$\sigma$ upper limits on molecular gas fractions of dormant galaxies lie at least $\sim0.3-2$\,dex below those expected for main-sequence galaxies, indicating low molecular gas reservoirs. 
Second, we did not find a clear systematic difference in molecular gas fractions between dormant galaxies and quiescent galaxies, given that most of these measurements are upper limits\footnote{We also performed a two-sample survival analysis accounting for the censored measurements. A log-rank test does not reveal a statistically significant difference between the dormant and quiescent populations ($p\simeq0.16$). However, the statistical power of this comparison is limited, as all but one of the measurements are upper limits.}.
Taken together, we find no evidence that dormant galaxies retain systematically larger molecular gas reservoirs than quiescent galaxies, despite being observed shortly after the onset of quenching.

\section{Sleeping or Dead? The fate of dormant galaxies}\label{Sec: discussion}
\subsection{Gas depletion or suppressed star formation efficiency?}

A rapid decline in star formation can arise through two broadly different scenarios: a reduction in the available molecular gas reservoir through gas depletion or removal, or a suppression of the efficiency with which the existing molecular gas is converted into stars. Distinguishing between these scenarios requires direct constraints on the molecular gas content. In the former case, recently quenched galaxies are expected to already exhibit reduced molecular gas reservoirs \citep[e.g.,][]{Dekel2006, Tacchella2016b}, whereas in the latter case, substantial molecular gas reservoirs may remain despite their strongly suppressed current star formation \citep[e.g.,][]{Martig2009, Xiao2018, Gensior2020}.

The key result of this work is that dormant galaxies already exhibit molecular gas fractions $\gtrsim$$0.3-2$\,dex below those of main sequence galaxies of similar mass, despite being observed shortly after a rapid decline in star formation. These fractions are consistent with those of quiescent galaxies given the current observational limits. 
These results favor a scenario in which the rapid shutdown of star formation is associated with depletion or removal of the molecular gas reservoir. Possible mechanisms include rapid gas consumption during an earlier episode of intense star formation, feedback-driven removal or heating of the cold gas, and suppression of further gas accretion. Our measurements do not distinguish among these processes, but they indicate that a substantial main-sequence-like molecular gas reservoir is no longer present in most of the well-constrained dormant galaxies.

In contrast, if the decline in star formation were primarily driven by a reduction in star formation efficiency, dormant galaxies could retain substantial molecular gas reservoirs while forming stars inefficiently, for example because the gas is dynamically stabilized or otherwise prevented from collapsing. Such a scenario would lead to relatively high molecular gas fractions together with long depletion timescales. However, the comparable gas fractions of dormant and quiescent galaxies do not favor suppressed star formation efficiency as the primary explanation for the majority of our sample. This mechanism may still contribute in individual objects, particularly among the $\sim30\%$ of the sample for which the current ALMA data are too shallow to provide meaningful constraints.

The similarity in molecular gas content between dormant and quiescent galaxies suggests that the transition from active star formation to a gas-poor state can occur rapidly. Rather than representing galaxies whose star formation has simply paused while retaining a substantial molecular gas reservoir, the dormant galaxies appear to have already reached a gas-poor state. Whether this state is temporary or represents the beginning of a longer transition toward quiescence depends critically on whether their gas reservoirs can be replenished, as discussed below.

\subsection{Bursty star formation and the subsequent fate of dormant galaxies} 

Recent observations and simulations have provided growing evidence for bursty star formation in high-redshift galaxies \citep[e.g.,][]{Sun2023_burstysf, Ciesla2024, Dome2024, Looser2024Natur, Faisst2025, McClymont2025, Xiao2026_massivesample}. The so-called ``mini-quenched'' or "napping" galaxies have been suggested to represent temporary low-SFR phases within such burst-quench cycles \citep[e.g.,][]{Looser2024Natur, Dome2024, Faisst2025, Baker2025_miniquench, Baker2026_twominiquench}. For example, \citet{Dome2024} find mini-quenching phases lasting $\sim20-40$\,Myr in galaxies with $\log(M_{\star}/M_\odot)=7-9$  at $z = 4-8$. However, this scenario has been studied primarily in lower-mass galaxies ($\log(M_{\star}/M_\odot)\lesssim9-9.5$), where stellar feedback can efficiently regulate or remove the cold gas supply. Our sample probes a higher mass range, with $\log(M_{\star}/M_\odot)>9.5$ and most dormant galaxies having $\log(M_{\star}/M_\odot)>10$ (Table~\ref{table: stellar}). At these higher masses, stellar feedback may be less efficient at removing gas from the deeper gravitational potential, while simulations suggest that AGN feedback can play an important role in regulating star formation and the gas reservoir \citep[e.g.,][]{Mutch2013, Hartley2023, Kurinchi-Vendhan2024, Sanati2025, Whitaker2026_REVIEW}. It therefore remains unclear whether the same burst-quench cycles established for low-mass galaxies also apply to the more massive dormant galaxies studied here.

Importantly, the low molecular gas fractions found in our work do not necessarily rule out such a burst-quench scenario. A burst of star formation can be followed by rapid gas consumption and feedback-driven gas removal, leaving a low gas reservoir during the subsequent low-SFR phase \citep[e.g.,][]{Dome2024, Sun2023_burstysf,McClymont2025, Cenci2026}. For example, \citet{McClymont2025} find a typical delay of only $\sim$21\,Myr between a star-formation burst and the subsequent rapid removal of gas from the ISM. If our dormant galaxies represent such a temporary low-SFR phase within a burst-quench cycle, their low molecular gas fractions could therefore be consistent with a post-burst, gas-poor state. Renewed star formation would then require re-accretion of previously removed gas or fresh gas accretion.

On the other hand, the higher stellar masses of our sample may point to a different subsequent evolution. Rapid gas consumption and AGN-driven gas removal have both been suggested as possible processes responsible for the rapid quenching of massive high-redshift galaxies \citep[e.g.,][]{Hartley2023, Kurinchi-Vendhan2024, Whitaker2026_REVIEW, Xiao2026_massivesample}. In this case, the low gas fractions observed here may instead mark a transition toward longer-lasting quiescence. Our current observations cannot distinguish whether the gas-poor state of dormant galaxies is a temporary phase within a burst-quench cycle or the beginning of a longer transition toward quiescence.

Distinguishing between these possibilities requires quantitative predictions for the evolution of the cold-gas reservoir during and after a burst, particularly at the stellar masses and redshifts probed by our sample. Current simulations already predict rapid changes in the ISM associated with bursty star formation \citep[e.g.,][]{Cenci2024, Dome2024, McClymont2025}, but predictions that can be directly compared with molecular gas fractions of massive dormant galaxies remain limited. Such predictions would allow the molecular gas measurements presented here to directly test whether these galaxies are consistent with a burst-quench cycle and, if so, how rapidly their gas reservoirs must be rebuilt before star formation can resume. Our results therefore neither rule out nor by themselves favor a bursty interpretation. They show, however, that any future rejuvenation would require substantial gas replenishment \citep[e.g.,][]{Remus2025}. In the absence of such replenishment, the similarity in molecular gas content between dormant and quiescent galaxies favors continued evolution toward quiescence, making these systems more consistent with becoming ``\textit{dead}'' than remaining temporarily ``\textit{sleeping}''.

\subsection{Caveats}\label{Sec: Caveats}

Here we list several caveats that should be considered when interpreting these results. First, all public ALMA measurements in our sample are upper limits, except for RUBIES-UDS-QG-z7, and the archival observations have highly heterogeneous sensitivities. Approximately 30\% of the sample has limits that lie on or above the expected main-sequence gas fractions and therefore do not yet provide meaningful constraints on the molecular gas reservoir. 
Moreover, an important open question is whether dormant galaxies have different gas depletion timescales from more evolved quiescent galaxies. However, the current upper limits on their molecular gas content prevent us from robustly constraining and comparing these timescales. Deeper and more homogeneous ALMA observations are therefore required to address this question in future observing cycles.

Second, the inferred molecular gas masses depend on the adopted line-to-gas conversion factors. For \cii, we use the empirical conversion factor of \citet{zanella2018}, although \cii\ emission can originate from multiple phases of the interstellar medium and the calibration has an intrinsic scatter of 0.3\,dex. Nevertheless, recent hydrodynamical simulations find that the global \cii-$M_{\rm mol}$ relation remains relatively robust across different evolutionary stages, including merger-induced starburst and quiescent phases \citep{miranda2026ciiglobalcoldgas}. For CO, the inferred molecular gas masses depend on both the assumed excitation ladder and $\alpha_{\rm CO}$. As described in Sect.~\ref{Sec: gas}, the metallicities expected at the typical stellar mass of our sample are consistent with our adopted Milky Way $\alpha_{\rm CO}$ value. These estimates, however, rely on mass--metallicity relations derived for star-forming galaxies \citep[e.g.,][]{Sharda2026,Faisst2026}. Among star-forming galaxies, lower-SFR systems generally exhibit higher gas-phase metallicities at fixed stellar mass, although this dependence weakens at the high-mass end \citep[e.g.,][]{Mannucci2010,Curti2020}. If this trend extends to our quiescent and dormant galaxies, their metallicities may be higher than inferred from the star-forming relations, further supporting our adopted $\alpha_{\rm CO}$. 
Nevertheless, uncertainties in individual gas-phase metallicities and CO excitation affect the CO-based gas masses and gas fractions. Using the same prescriptions as recent studies of high-redshift quiescent galaxies reduces systematic differences in the comparison (Fig.~\ref{fgas}).

Finally, the sample is selected from publicly available JWST/NIRSpec prism observations with archival ALMA coverage and is therefore neither complete nor selected according to uniform ALMA sensitivity. The distinction between dormant and quiescent galaxies also depends on the inferred recent SFH and the adopted spectral criteria. Nevertheless, the consistent comparison within our sample, together with literature measurements of quiescent galaxies, supports the conclusion that recently quenched dormant galaxies do not generally retain molecular gas reservoirs substantially larger than those of quiescent systems.

\section{Conclusions}\label{Sec: conclusion}

In this work, we investigate the molecular gas content of recently quenched ``dormant'' galaxies at $z>3$, to determine whether their recent decline in star formation is mainly dominated by the depletion/removal of their molecular gas reservoirs or the suppressed star formation efficiency. We combine all JWST/NIRSpec prism spectroscopy and archival ALMA observations of \cii~158\,$\mu$m and CO lines ($1\leq J_{\rm up}\leq6$) publicly available to date.
Our main conclusions are as follows:

\begin{itemize}

    \item \textbf{Recent and rapid quenching.} Dormant galaxies, in general, show clear evidence for a recent and rapid decline in star formation. While most of them remain close to the SFMS when considering the SFR averaged over the past 100\,Myr, they shift systematically toward lower SFRs on the 10\,Myr timescale. They also preferentially exhibit lower $\mathrm{SFR}_{10\,\mathrm{Myr}}/\mathrm{SFR}_{100\,\mathrm{Myr}}$ ratios than both quiescent galaxies and star-forming galaxies, indicating that they are caught shortly after the onset of quenching.

    \item \textbf{Low molecular gas fractions.} The molecular gas fractions of dormant galaxies show no apparent systematic difference from those of quiescent galaxies, including both the quiescent galaxies in our sample and previously reported ones. Although the heterogeneous depths of the archival ALMA observations prevent meaningful constraints for $\sim$$30\%$ of the sample, the remaining $\sim$$70\%$ are sufficiently constrained to lie $\sim0.3-2$\,dex below the molecular gas fractions expected for main-sequence galaxies. 

    \item \textbf{Rapid quenching associated with gas depletion or removal.} The low molecular gas fractions of dormant galaxies shortly after the onset of quenching favor a scenario in which their rapid shutdown of star formation is associated with depletion or removal of the molecular gas reservoir, rather than primarily with suppressed SFE while retaining a substantial gas reservoir. Their molecular gas content is already comparable to that of quiescent galaxies, suggesting that the transition to a gas-poor state can occur rapidly.

\end{itemize}

Taken together, our results provide the first systematic constraints on the molecular gas reservoirs of recently quenched dormant galaxies at $z>3$, suggesting that these systems are already transitioning toward quiescence. Despite being observed shortly after the rapid decline in their star formation, their molecular gas reservoirs are already comparable to those of quiescent galaxies. Rather than systems whose star formation has merely paused while retaining a substantial molecular gas reservoir, they appear to have already reached a gas-poor state. Therefore, these galaxies are more consistent with becoming ``\textit{dead}'' than with being temporarily ``\textit{sleeping}'', if no substantial gas replenishment occurs in the future. 

However, these first constraints are based predominantly on upper limits from archival ALMA observations with heterogeneous sensitivities. Deeper and more homogeneous ALMA observations of a larger sample of dormant galaxies are therefore needed to establish whether rapid gas depletion is a general feature of the dormant phase and to constrain how rapidly the molecular gas reservoir is depleted or removed during the transition toward quiescence.

\begin{acknowledgements}
The data products presented herein were retrieved from the Dawn JWST Archive (DJA). DJA is an initiative of the Cosmic Dawn Center (DAWN), which is funded by the Danish National Research Foundation under grant DNRF140. 
This work has received funding from the Swiss State Secretariat for Education, Research and Innovation (SERI) under contract number MB22.00072, as well as from the Swiss National Science Foundation (SNSF) through project grant 200020\_207349. 
LB acknowledges support from the Sussex Astronomy Centre STFC Consolidated Grant 2023-2026 (ST/X001040/1).
YF was supported by JSPS KAKENHI Grant Numbers JP23K13149 and 26K00736.
LCK acknowledges support from UK Research and Innovation through a Future Leaders Fellowship, grant agreement MR/X035166/1.
The work presented in this paper is based on observations made
with the NASA/ESA/CSA JWST. The data were obtained from the
Mikulski Archive for Space Telescopes at the Space Telescope
Science Institute, which is operated by the Association of
Universities for Research in Astronomy, Inc., under NASA
contract NAS 5-03127 for JWST.
These observations are associated with the following programmes:
\#1180, \#1212, \#1214, \#1215, \#1286,  \#2198, \#2565,   \#4233 , and \#6368.
This paper makes use of the following ALMA data:
ADS/JAO.ALMA\#2013.1.01292.S,
ADS/JAO.ALMA\#2015.1.01205.S,
ADS/JAO.ALMA\#2015.A.00026.S,
ADS/JAO.ALMA\#2016.1.01001.S,
ADS/JAO.ALMA\#2018.1.01841.S,
ADS/JAO.ALMA\#2018.A.00037.S,
ADS/JAO.ALMA\#2021.1.00024.S,
ADS/JAO.ALMA\#2021.1.00225.S,
ADS/JAO.ALMA\#2021.1.01005.S,
ADS/JAO.ALMA\#2021.1.01291.S,
ADS/JAO.ALMA\#2022.1.01604.S,
ADS/JAO.ALMA\#2023.1.01016.S,
ADS/JAO.ALMA\#2023.1.01292.S,
ADS/JAO.ALMA\#2023.1.01296.S,
ADS/JAO.ALMA\#2024.1.01213.S,
ADS/JAO.ALMA\#2024.1.01662.S,
ADS/JAO.ALMA\#2024.1.01744.S,
and ADS/JAO.ALMA\#2025.1.00092.S.
ALMA is a partnership of ESO (representing its member states), NSF (USA) and NINS (Japan), together with NRC (Canada), NSTC and ASIAA (Taiwan), and KASI (Republic of Korea), in cooperation with the Republic of Chile. The Joint ALMA Observatory is operated by ESO, AUI/NRAO and NAOJ. The National Radio Astronomy Observatory and Green Bank Observatory are facilities of the U.S. National Science Foundation operated under cooperative agreement by Associated Universities, Inc.

\end{acknowledgements}

\bibliography{reference_thispaper}{}
\bibliographystyle{aa}

\onecolumn
\begin{appendix}

\section{JWST image and \texttt{Bagpipes} SED fitting of our sample. }\label{appendix_sed} 
\begin{figure*}[h!]
\centering
\includegraphics[width=14.3cm]{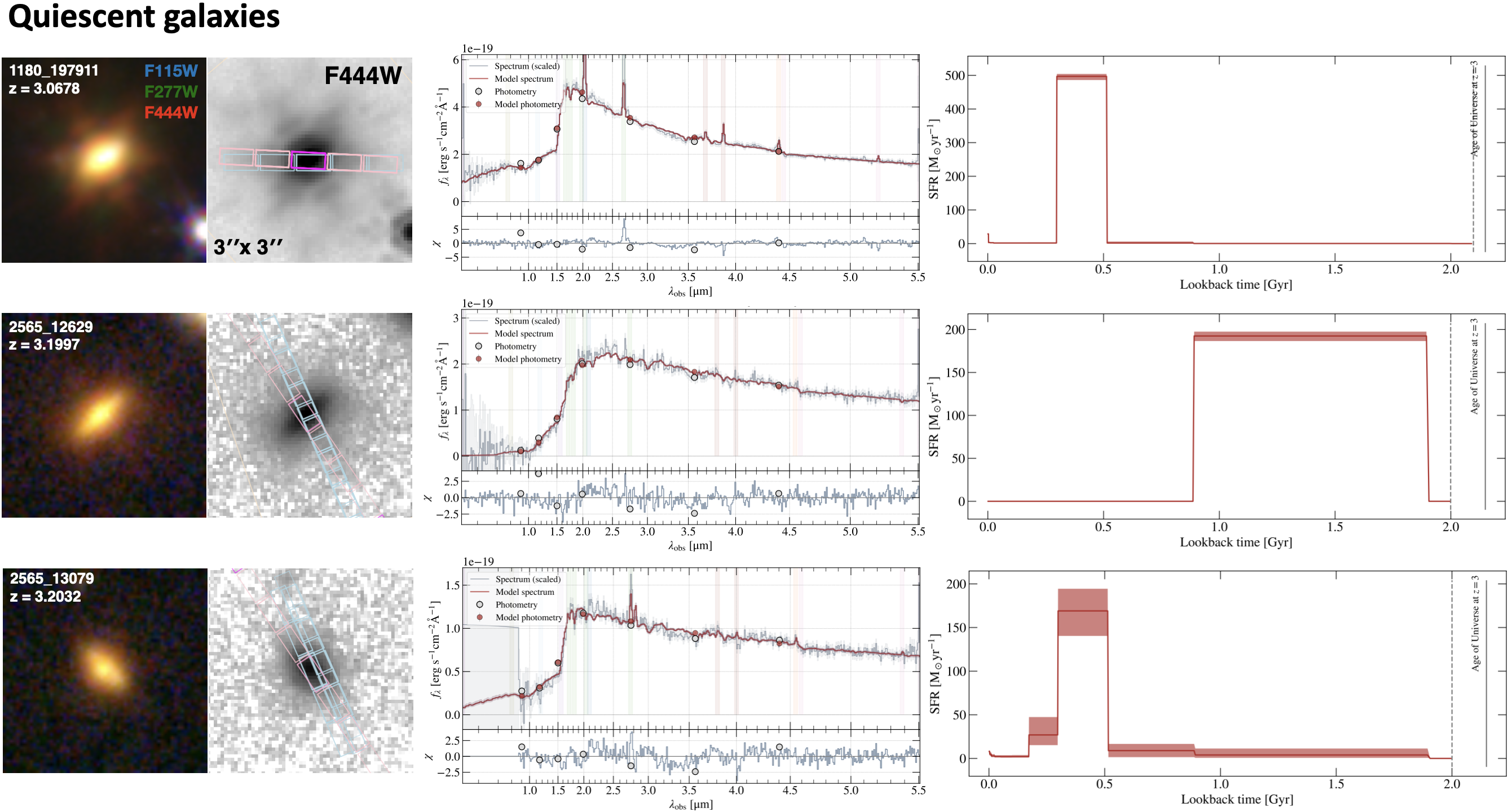} \hspace{-2mm}
\includegraphics[width=14.3cm]{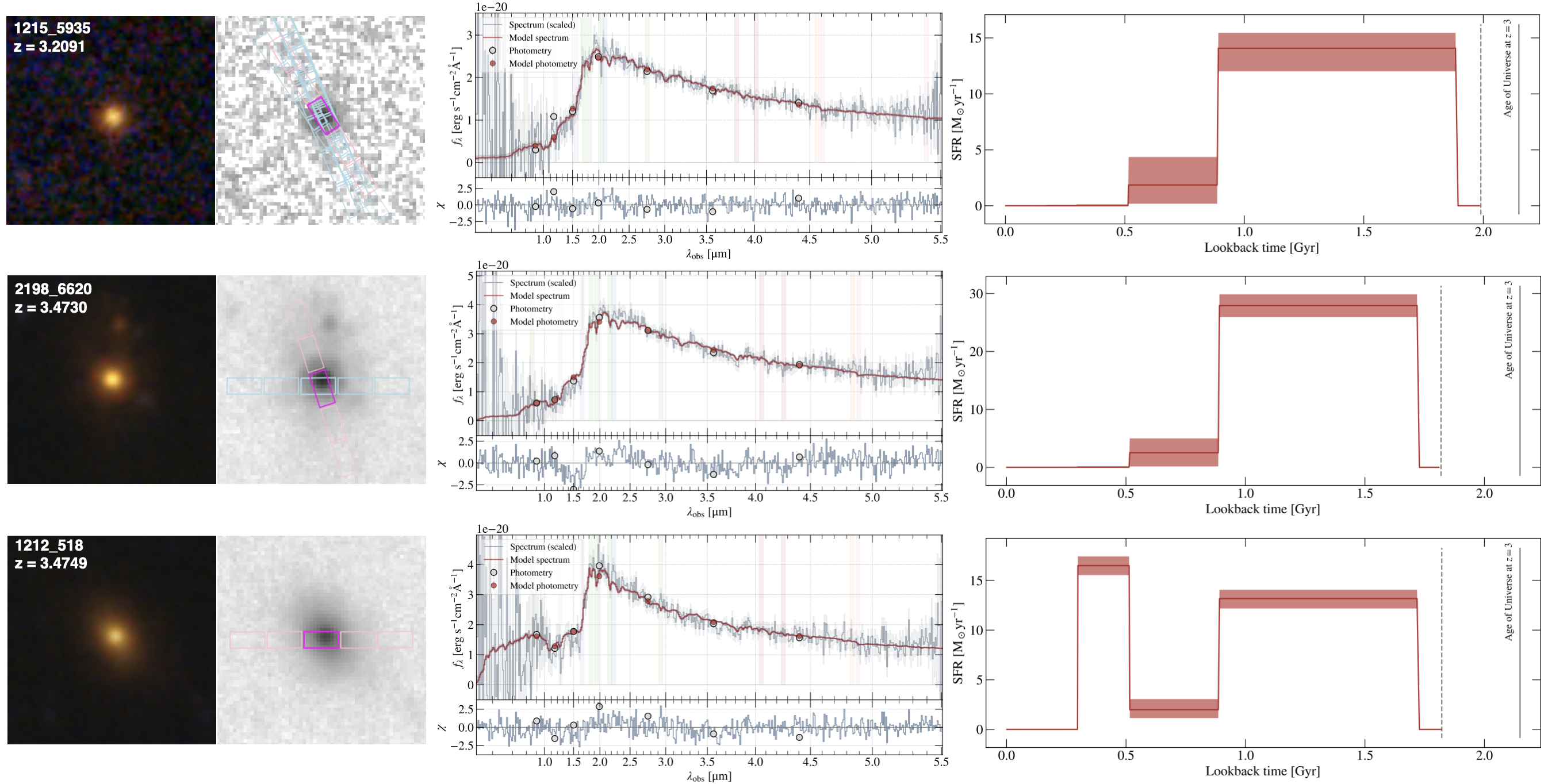} \hspace{-2mm}
\includegraphics[width=14.3cm]{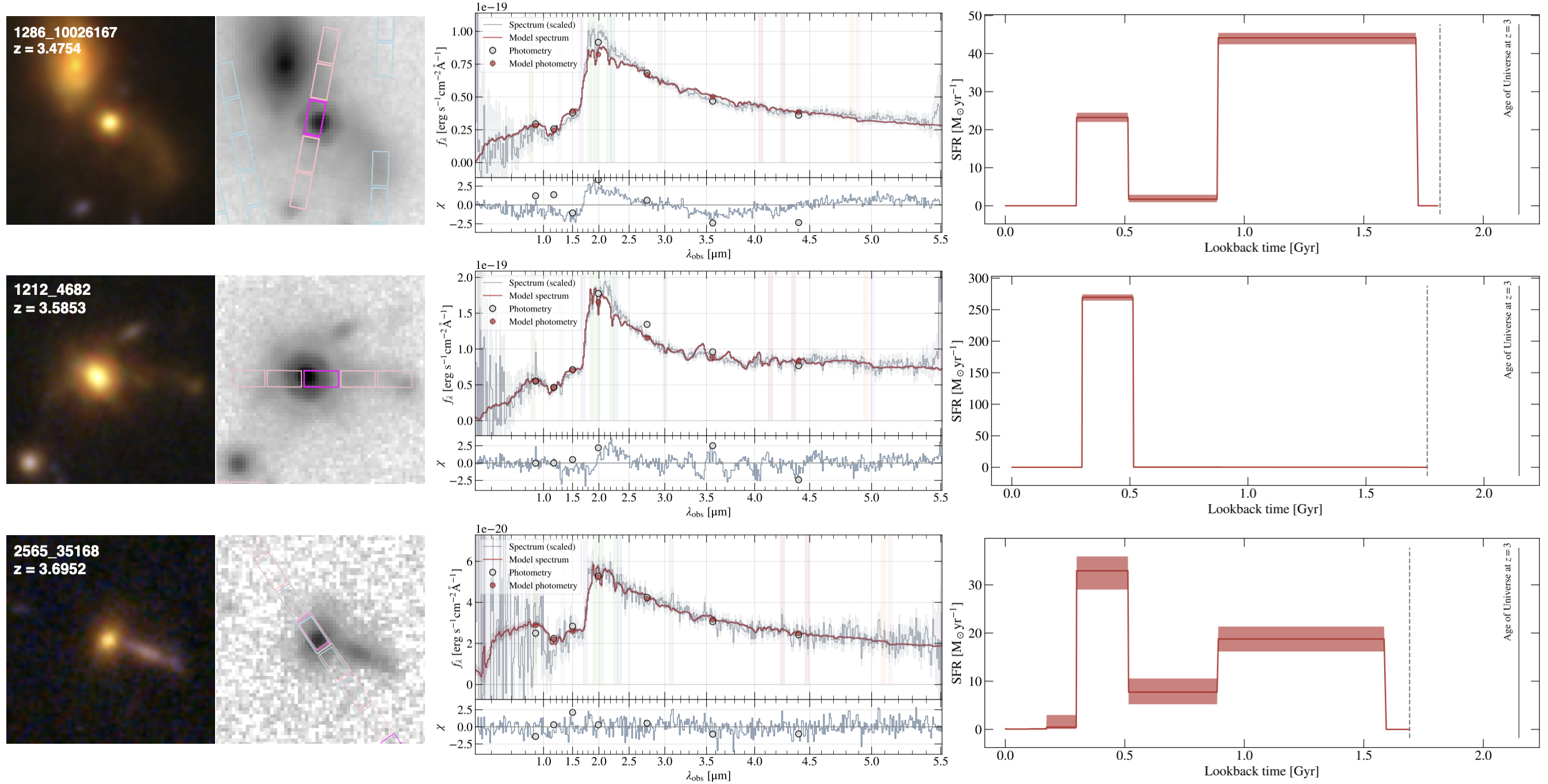}
\caption{\textbf{Joint SED fitting of photometry and prism spectra (see Sect.~\ref{Sect:parent}) of our sample of quiescent galaxies.} From left to right, we show the $3^{\prime\prime}\times 3^{\prime\prime}$ RGB image (F115W in blue, F277W in green, and F444W in red), the F444W image stamp, the best-fit \texttt{Bagpipes} model (red curve) together with the photometry (grey points) and prism spectrum (grey curve), and the inferred SFH \citep{gottumukkala2026}.
}
\label{fig:SEDQG}
\end{figure*}

\begin{figure*}[h!]
\centering
\includegraphics[width=14.3cm]{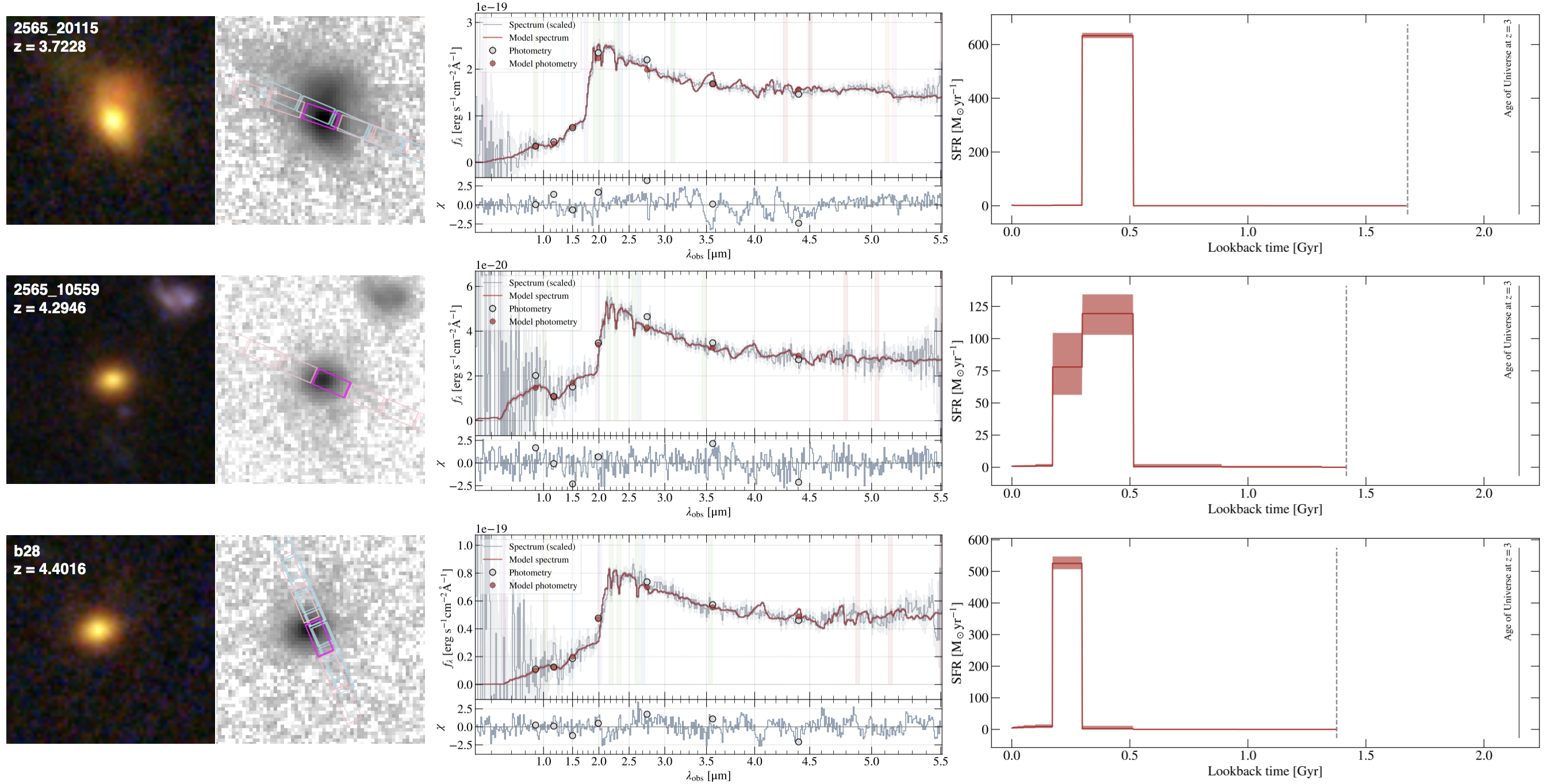} \hspace{-2mm}
\includegraphics[width=14.3cm]{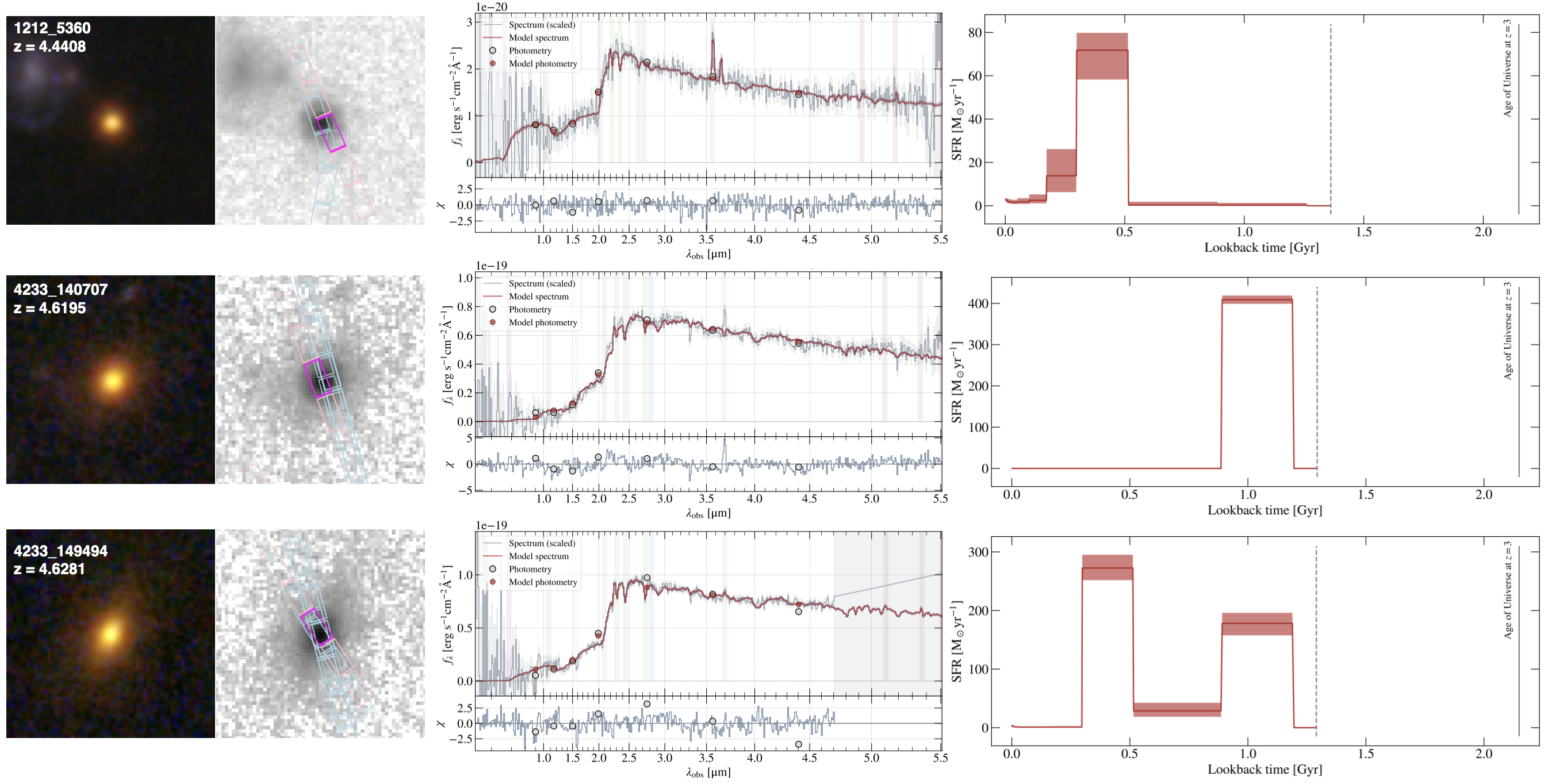} \hspace{-2mm}
\caption{\textbf{Joint SED fitting of photometry and prism spectra (see Sect.~\ref{Sect:parent}) of our sample of quiescent galaxies.} Continued. 
}
\label{fig:SEDQG2}
\end{figure*}
\begin{figure*}[h!]
\centering
\includegraphics[width=14.3cm]{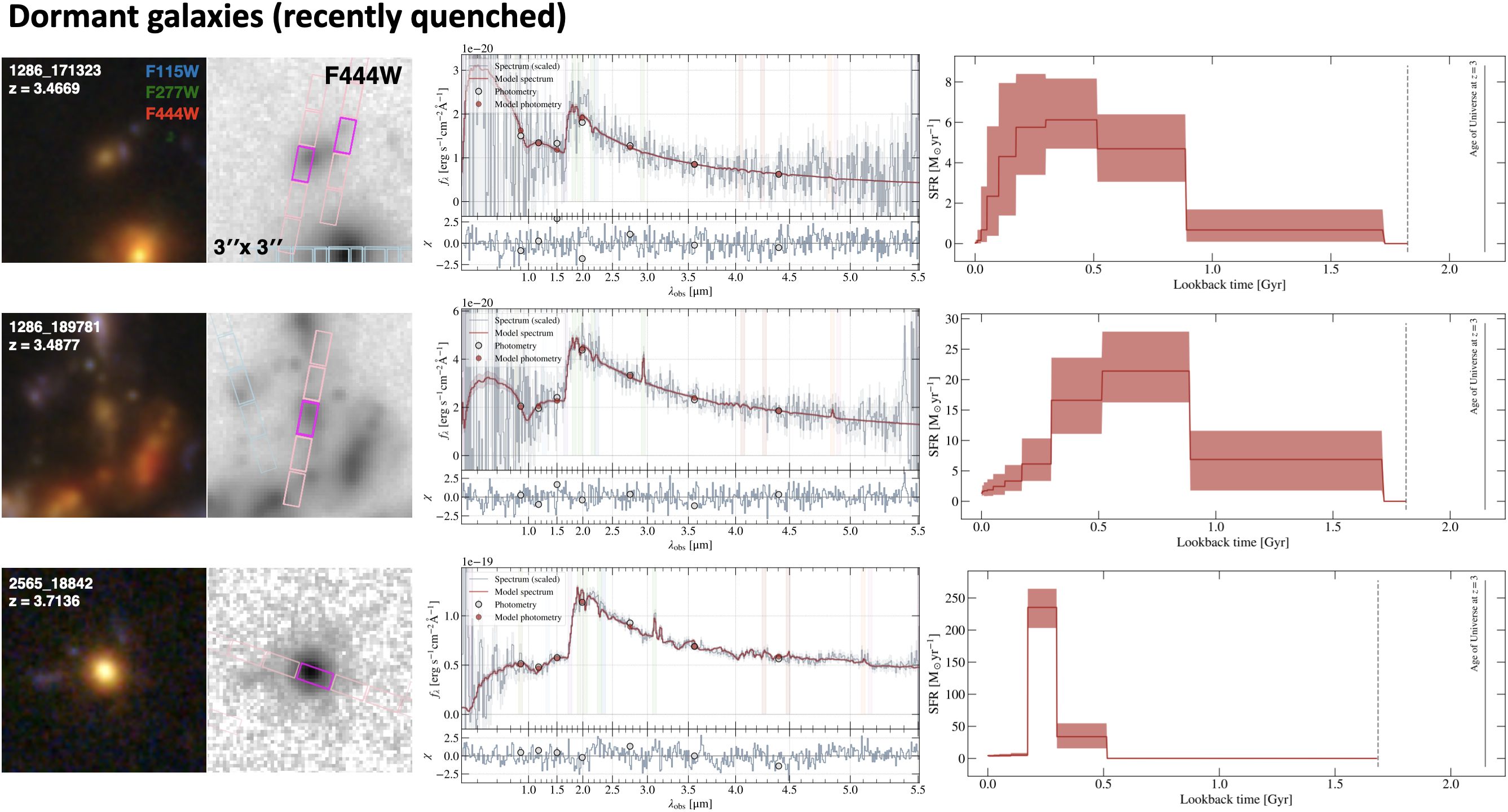} \hspace{-2mm}
\includegraphics[width=14.3cm]{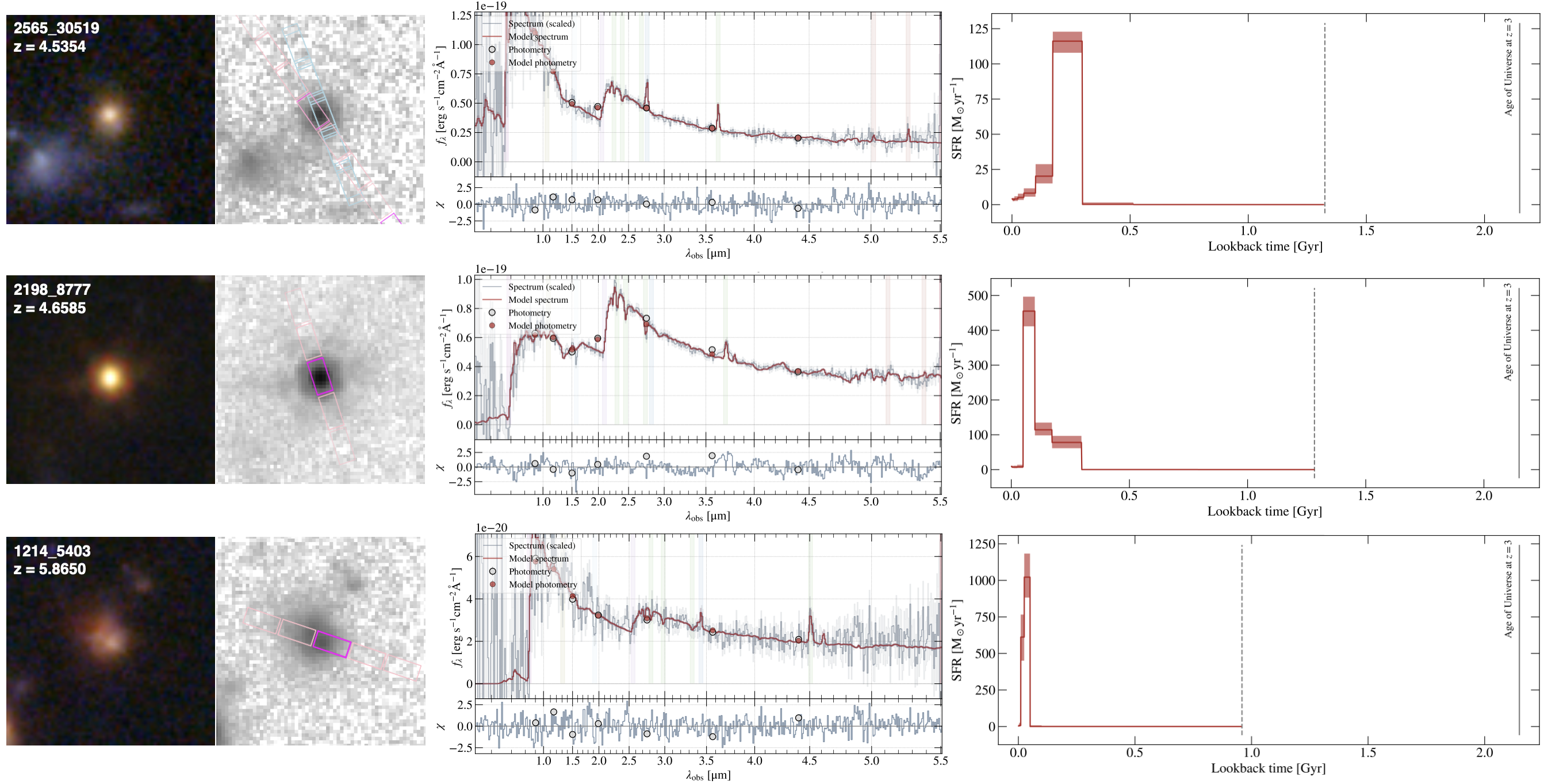} \hspace{-2mm}
\includegraphics[width=14.3cm]{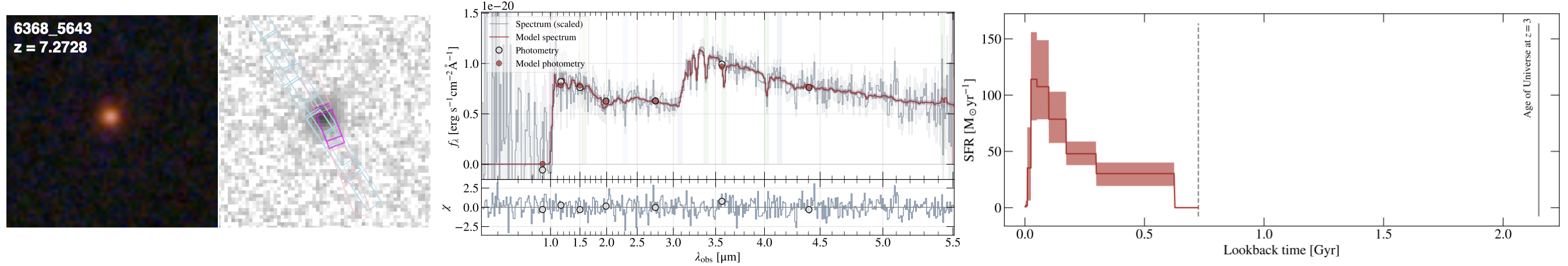}
\caption{\textbf{Joint SED fitting of photometry and prism spectra (see Sect.~\ref{Sect:parent}) of our sample of dormant galaxies.} From left to right, we show the $3^{\prime\prime}\times 3^{\prime\prime}$ RGB image (F115W in blue, F277W in green, and F444W in red), the F444W image stamp, the best-fit \texttt{Bagpipes} model (red curve) together with the photometry (grey points) and prism spectrum (grey curve), and the inferred SFH \citep{gottumukkala2026}.
}
\label{fig:SEDDG}
\end{figure*}

\clearpage
\section{ALMA observations of our sample. }\label{appendix_alma} 
\begin{figure*}[h!]
\centering
\includegraphics[width=14.3cm]{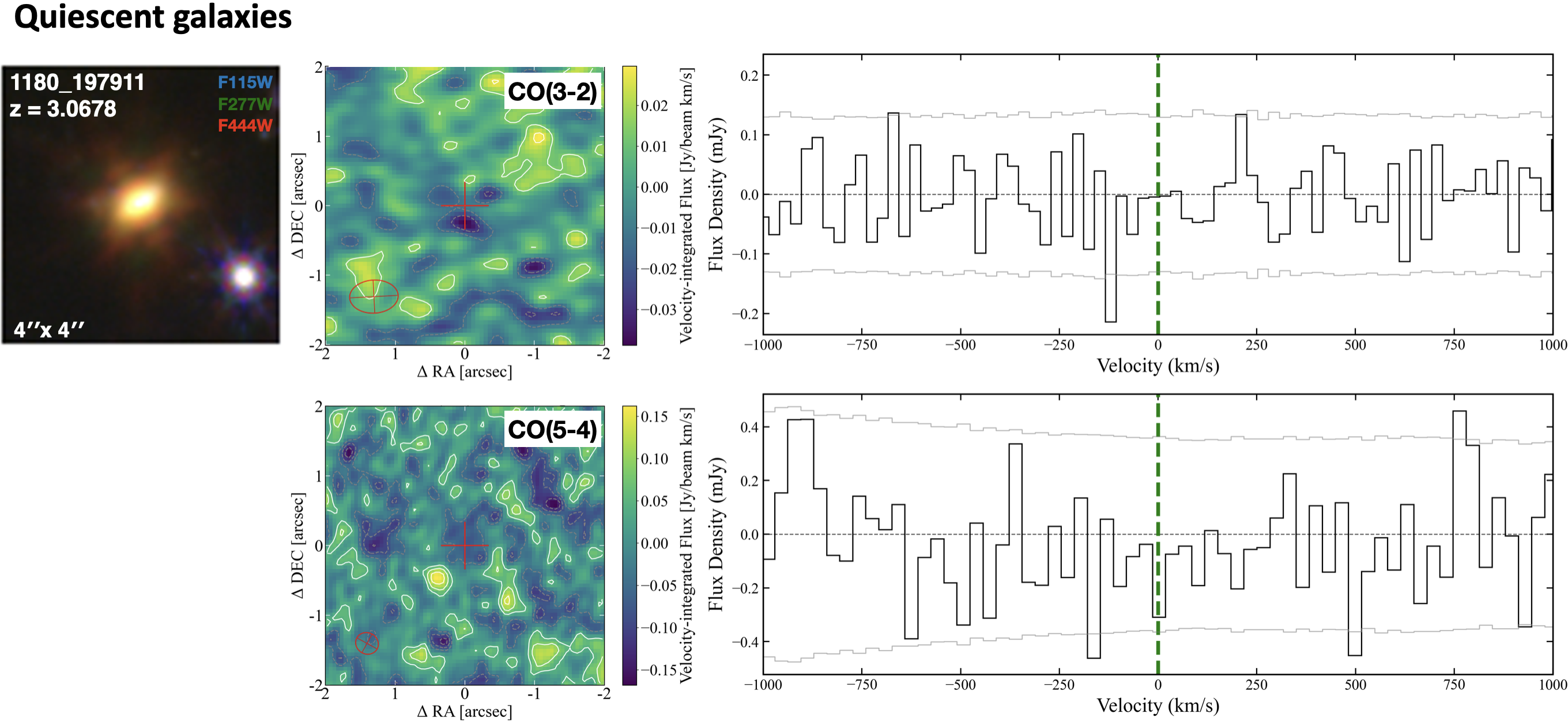} \hspace{-2mm}
\includegraphics[width=14.3cm]{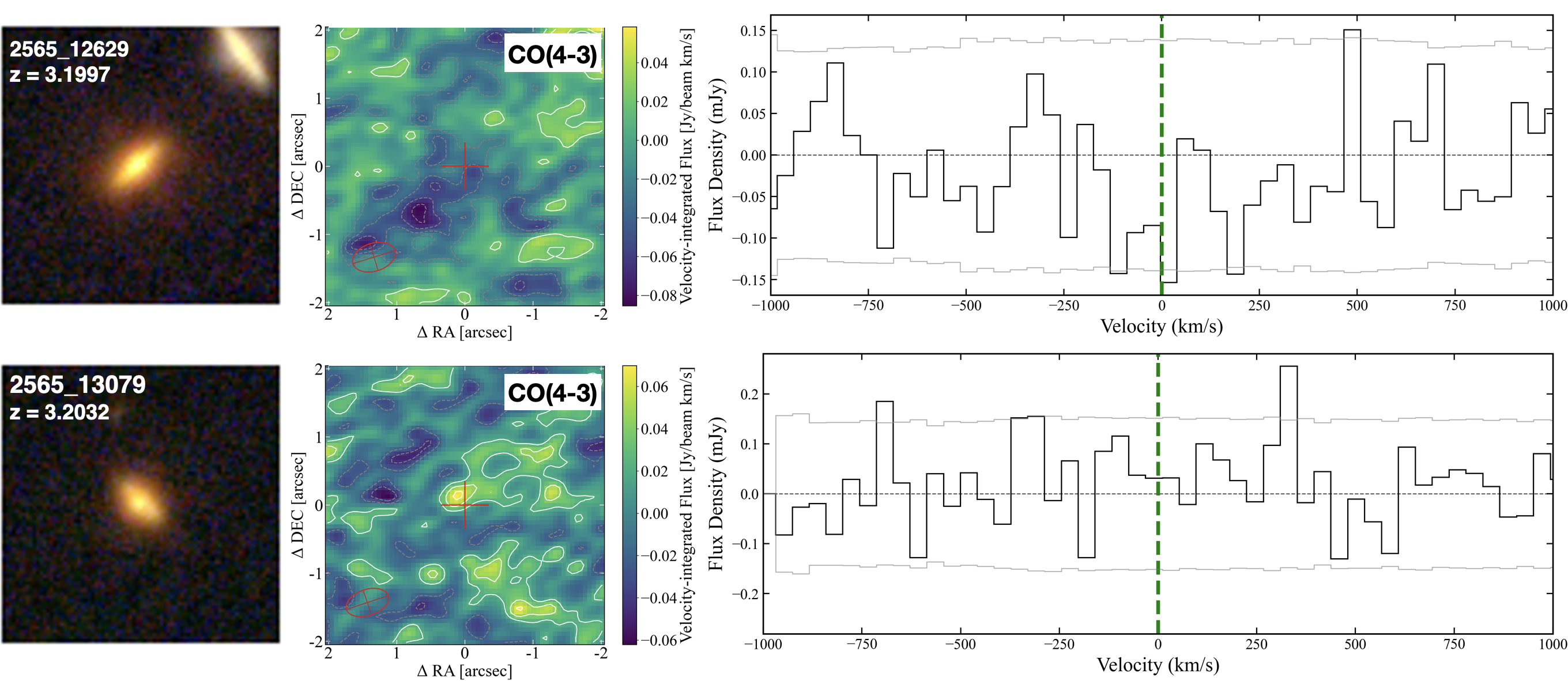} \hspace{-2mm}
\includegraphics[width=14.3cm]{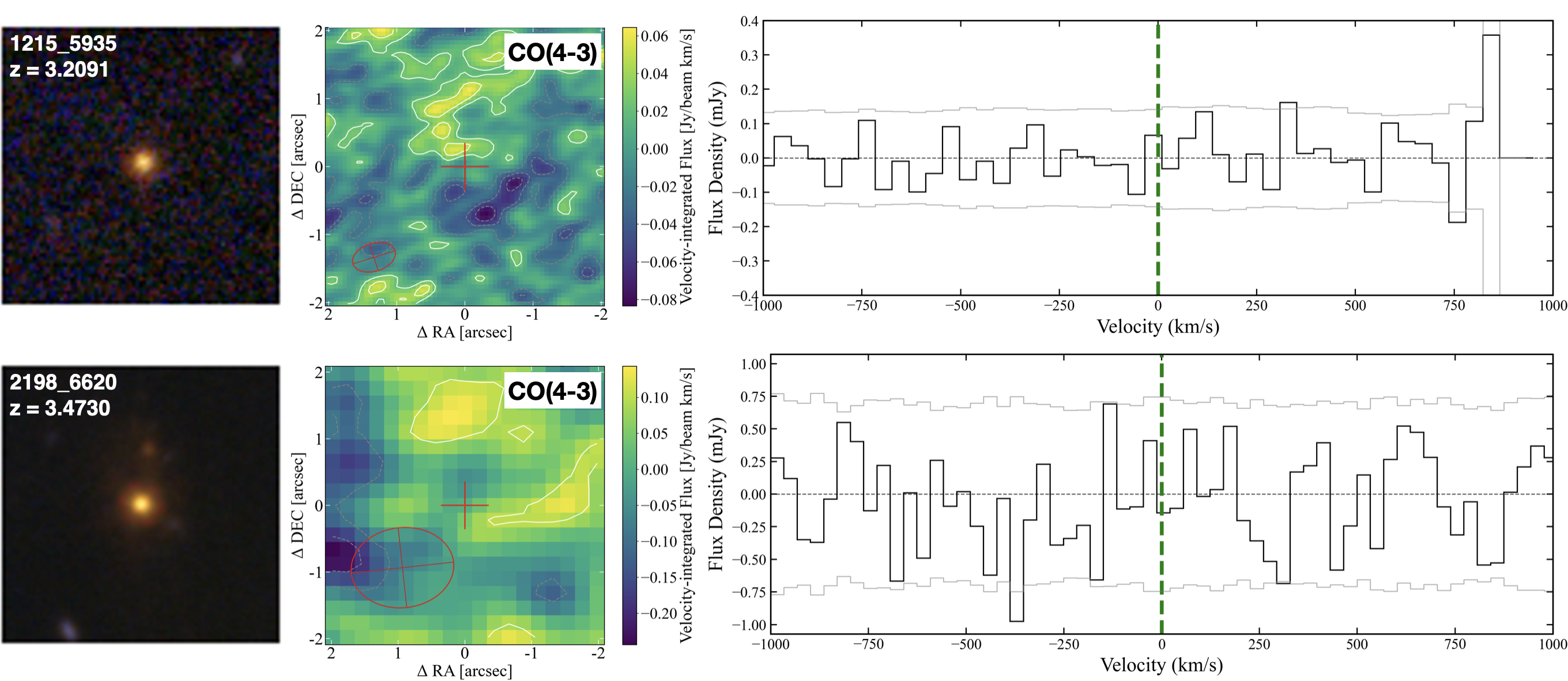} \hspace{-2mm}
\caption{\textbf{ALMA non-detections of \cii$\,158\,{\rm \mu m}$ and CO emission lines for the quiescent galaxies.} \textit{Left}: $4^{\prime\prime}\times4^{\prime\prime}$ JWST RGB images (F115W in blue, F277W in green, and F444W in red). \textit{Middle}: ALMA moment-0 maps of the \cii or CO lines (assuming a line width of 250 km s$^{-1}$; $4^{\prime\prime}\times4^{\prime\prime}$). The contour levels start at 1$\sigma$ and increase in steps of $\pm1\sigma$, where positive and negative contours are solid and dashed, respectively. The synthesized beam is shown in the lower-left corner. \textit{Right}: ALMA 1D spectra binned to 30-50 km s$^{-1}$. The grey solid lines show the associated 1$\sigma$ uncertainty. The vertical green dashed lines mark the locations of expected \cii$\,158\,{\rm \mu m}$ or CO lines based on the JWST spectroscopic redshifts. The spectra are extracted at the source positions using beam-sized apertures, assuming that the emission line is unresolved. 
      }
\label{fig:almaQG1}
\end{figure*}
\begin{figure*}[h!]
\centering
\includegraphics[width=14.3cm]{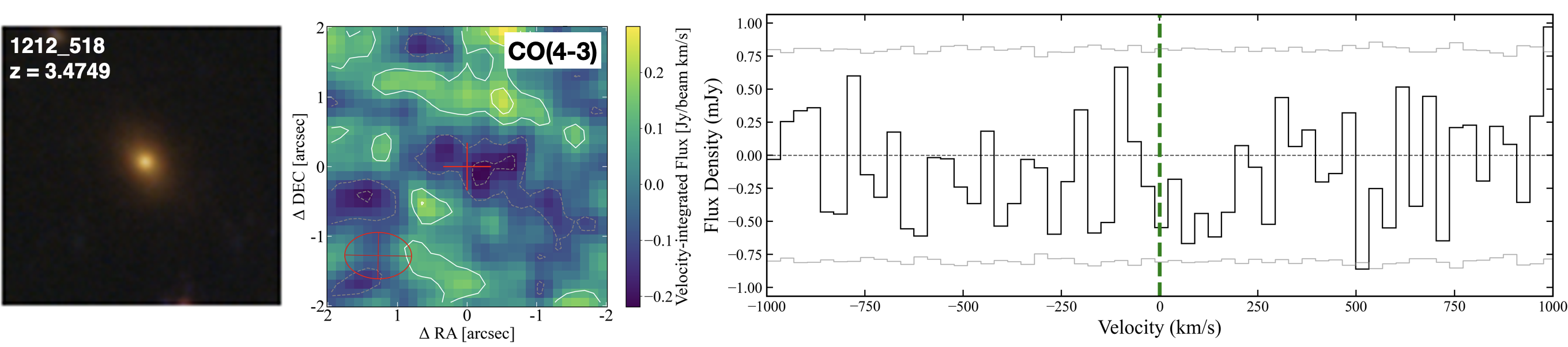} \hspace{-2mm}
\includegraphics[width=14.3cm]{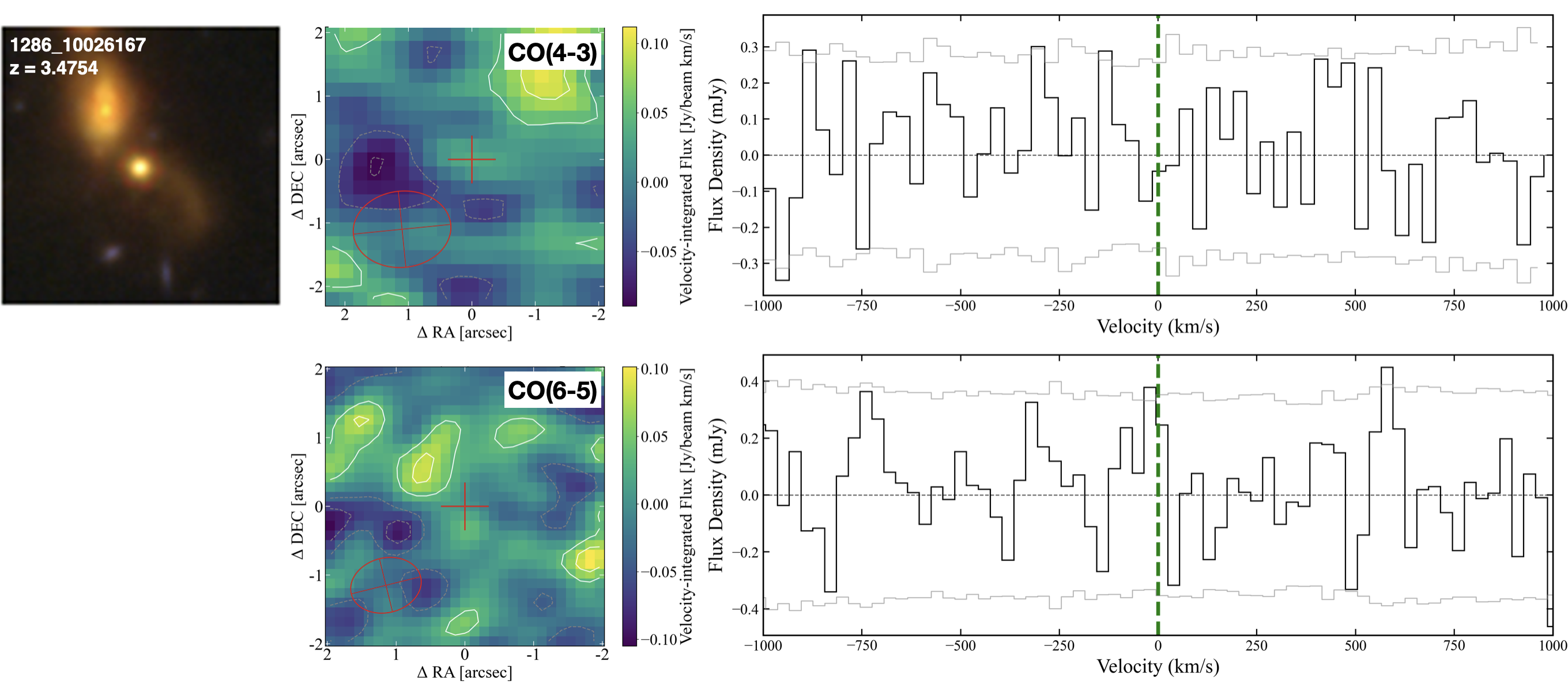} \hspace{-2mm}
\includegraphics[width=14.3cm]{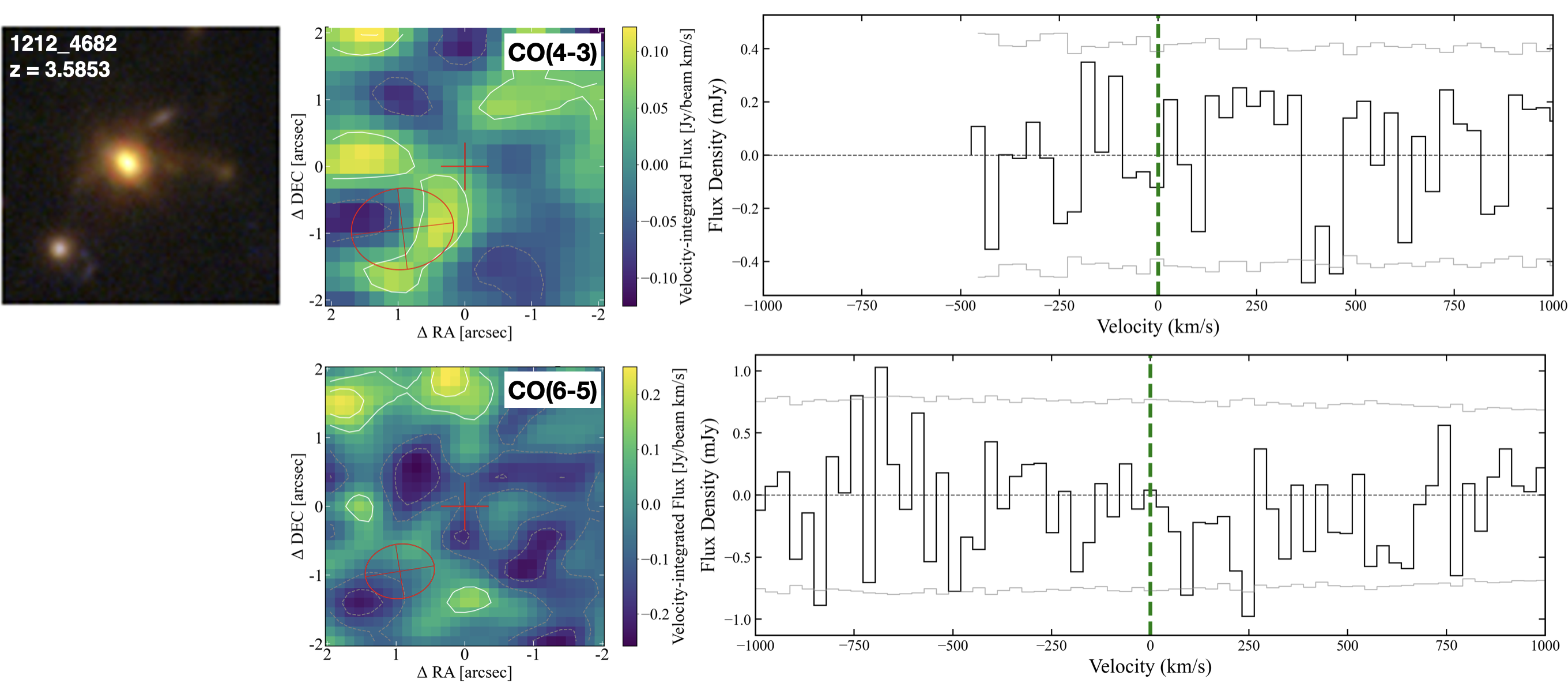} \hspace{-2mm}
\includegraphics[width=14.3cm]{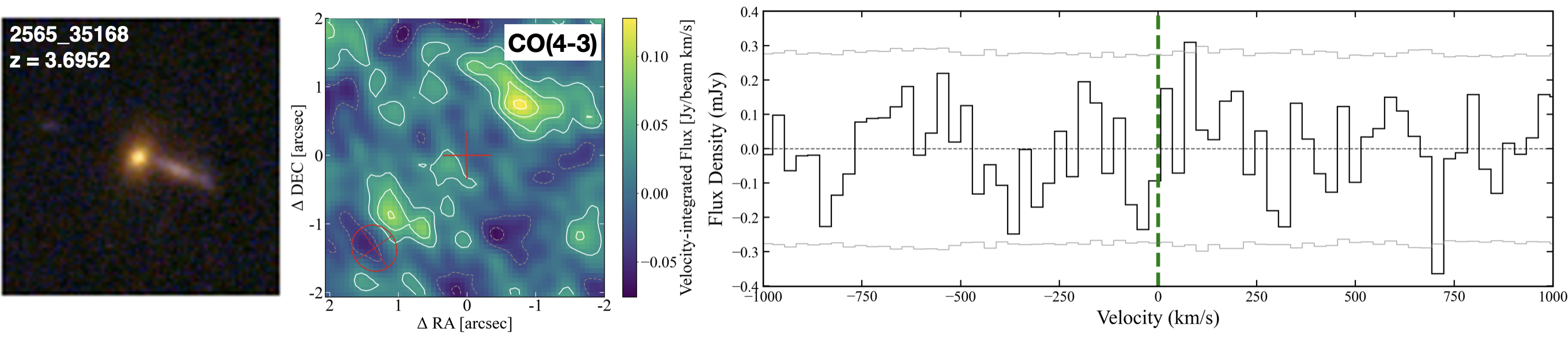} \hspace{-2mm}
\caption{Continued. 
}
\label{fig:almaQG2}
\end{figure*}

\begin{figure*}[h!]
\centering
\includegraphics[width=14.3cm]{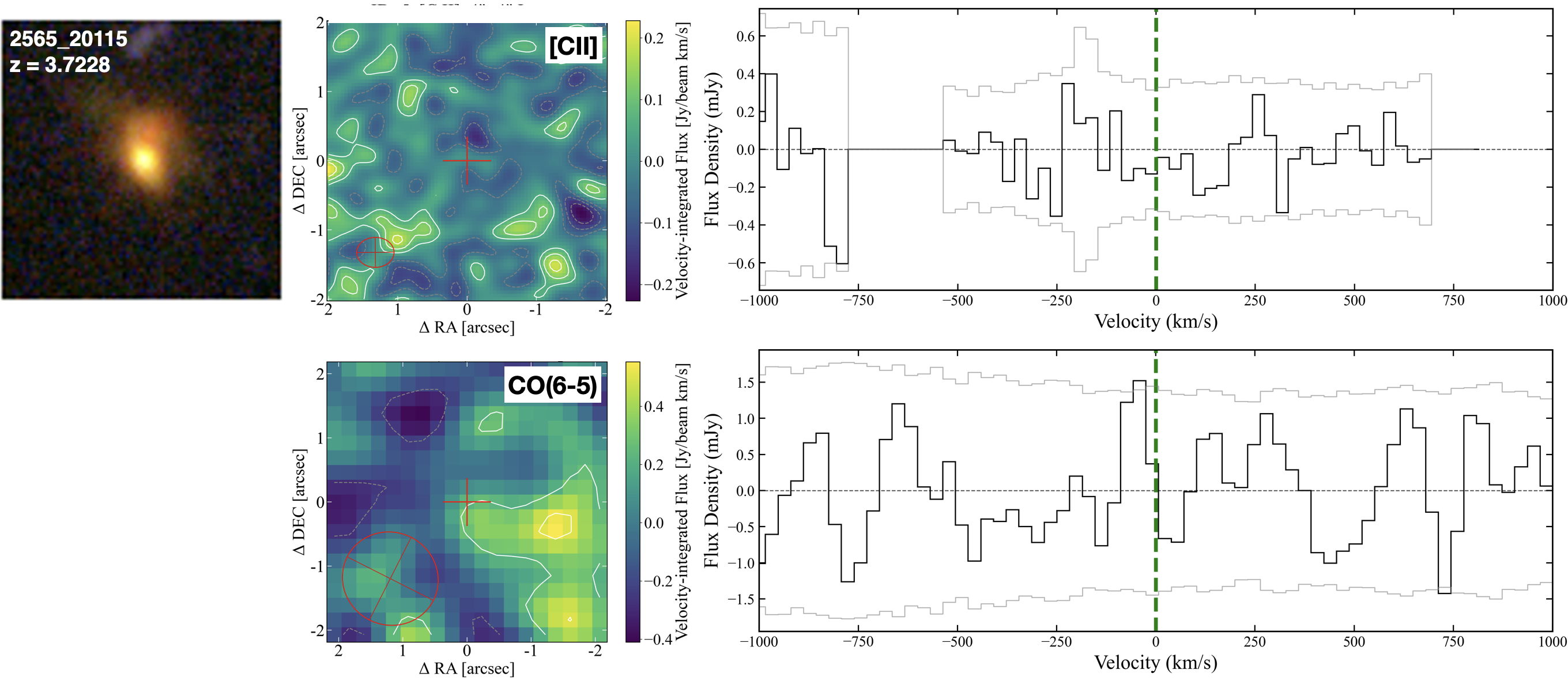} \hspace{-2mm}
\includegraphics[width=14.3cm]{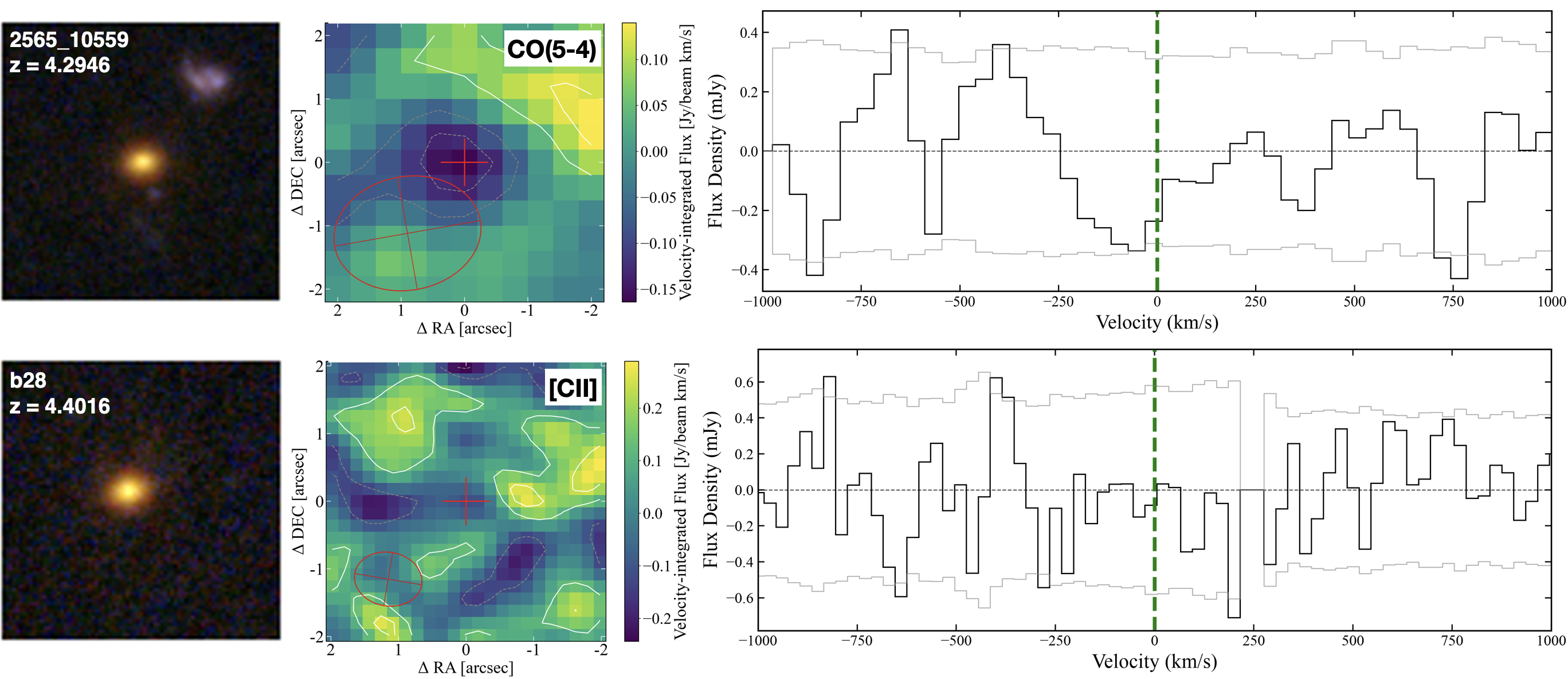} \hspace{-2mm}
\includegraphics[width=14.3cm]{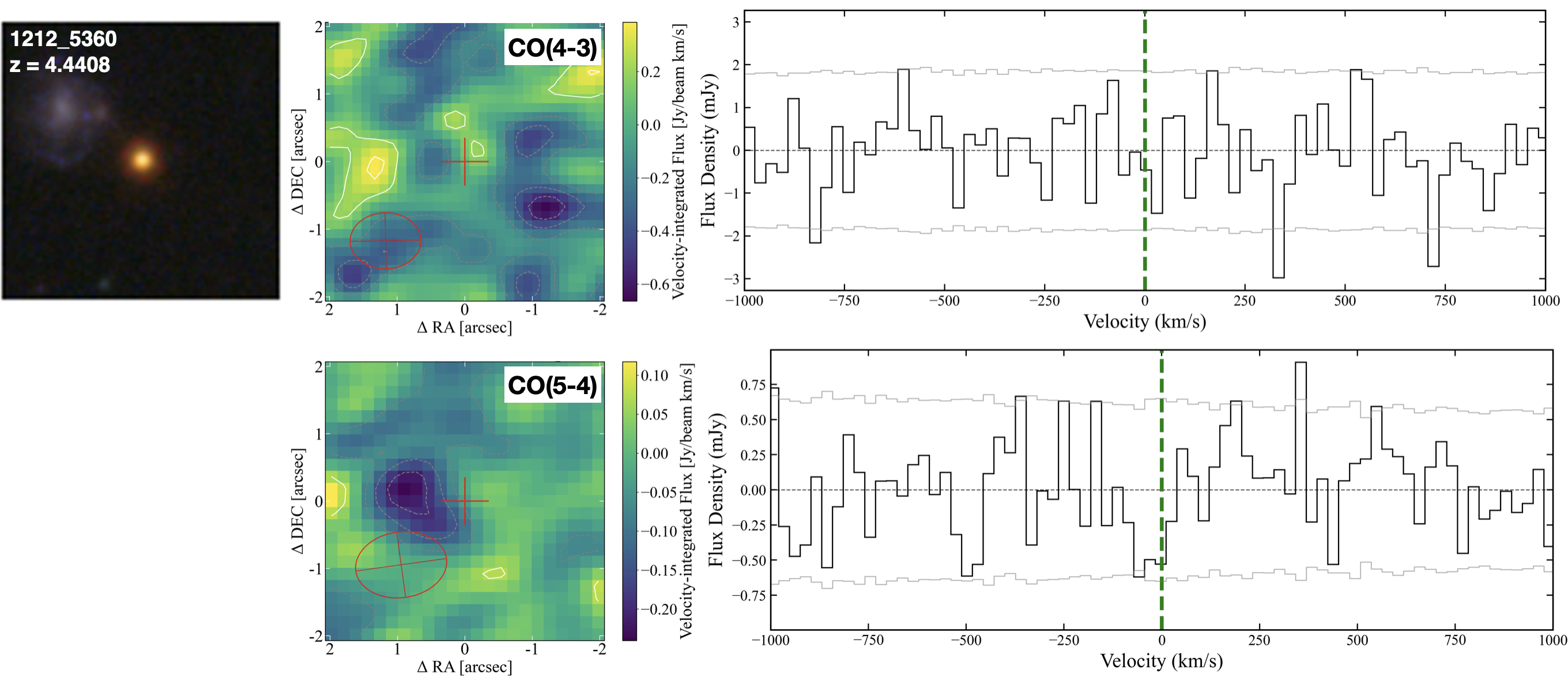} \hspace{-2mm}
\includegraphics[width=14.3cm]{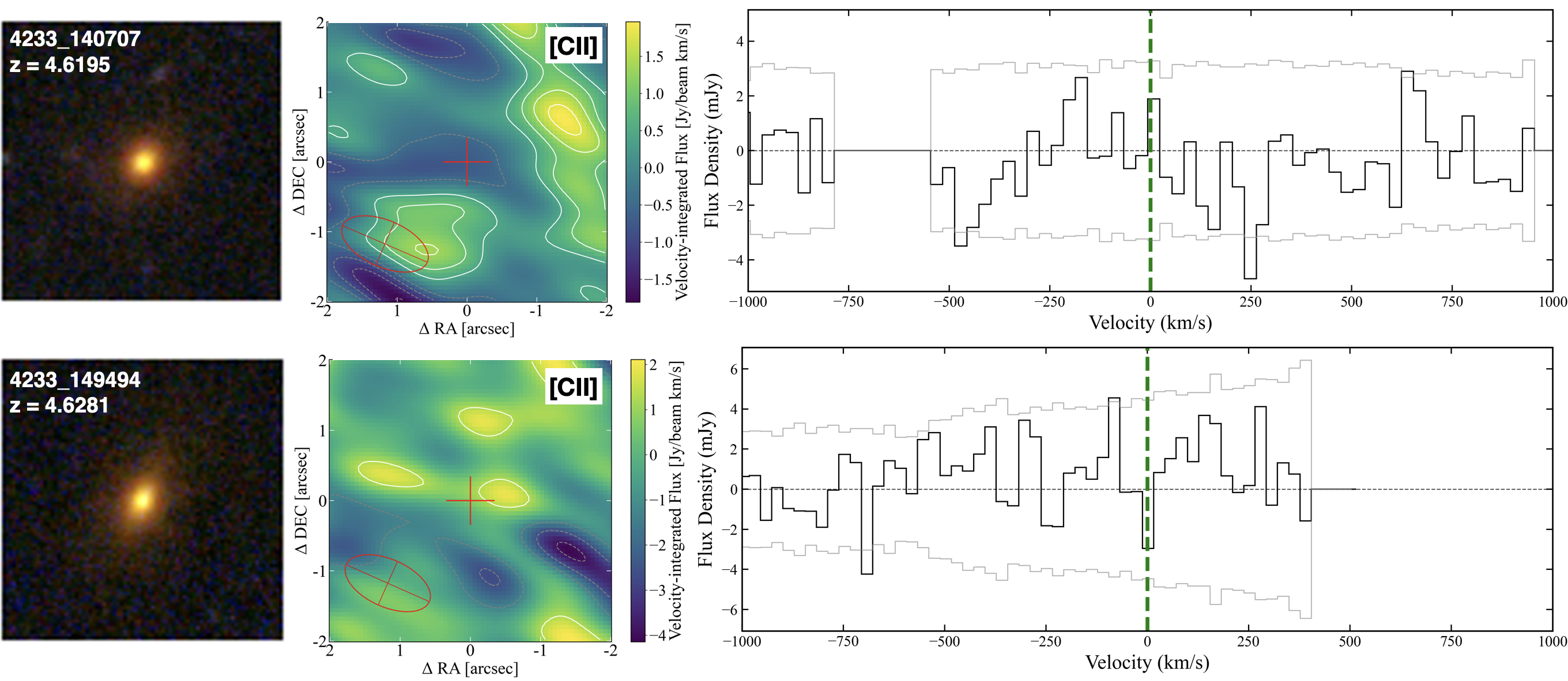} \hspace{-2mm}
\caption{Continued. 
}
\label{fig:almaQG3}
\end{figure*}

\begin{figure*}[h!]
\centering
\includegraphics[width=14.3cm]{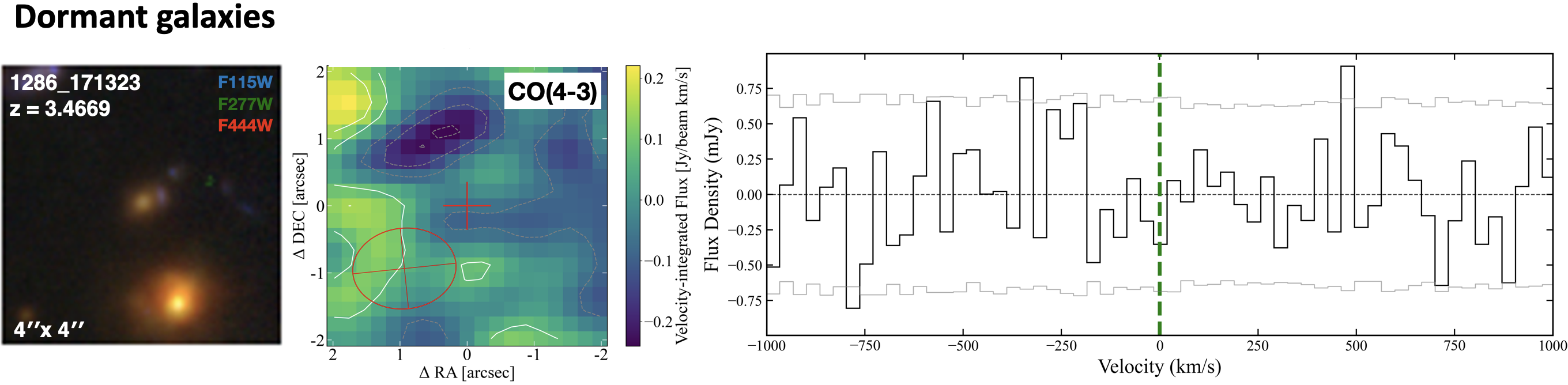} \hspace{-2mm}
\includegraphics[width=14.3cm]{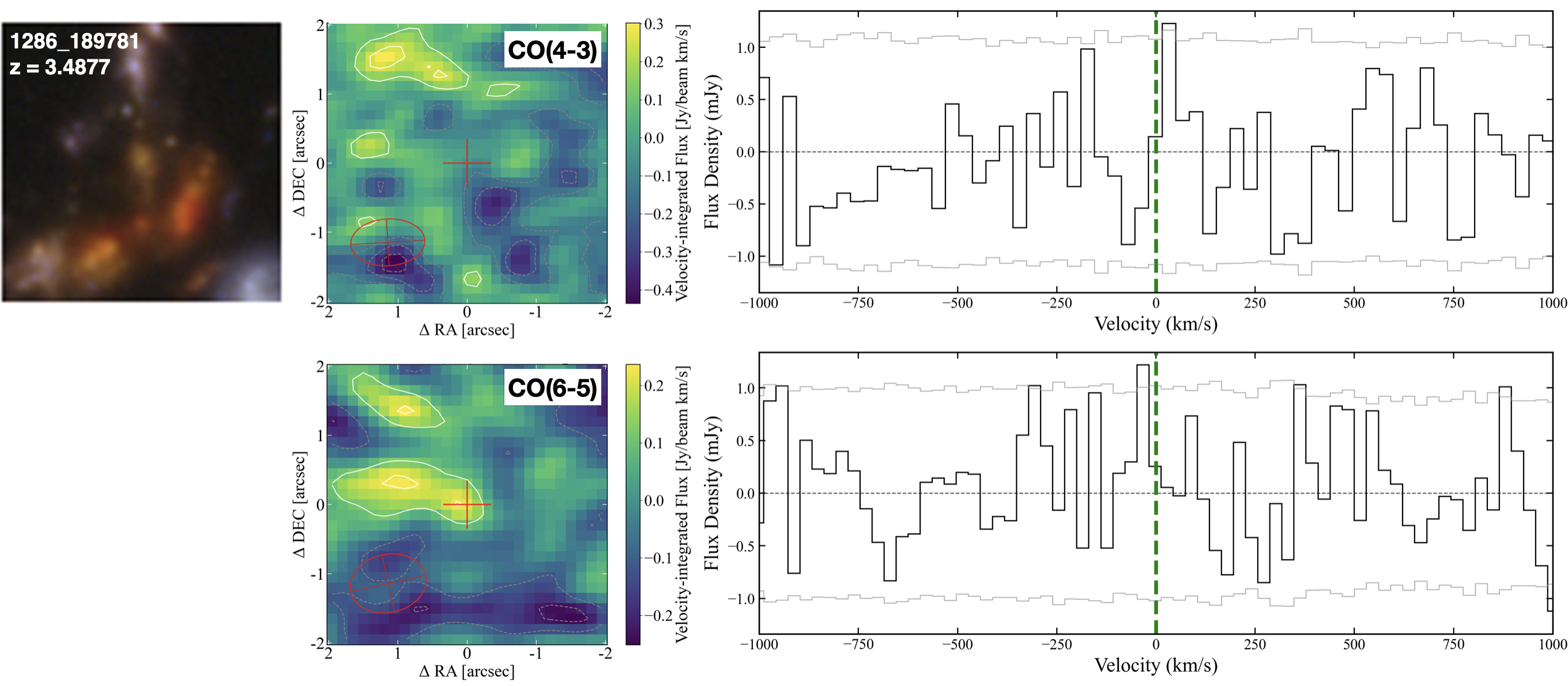} \hspace{-2mm}
\includegraphics[width=14.3cm]{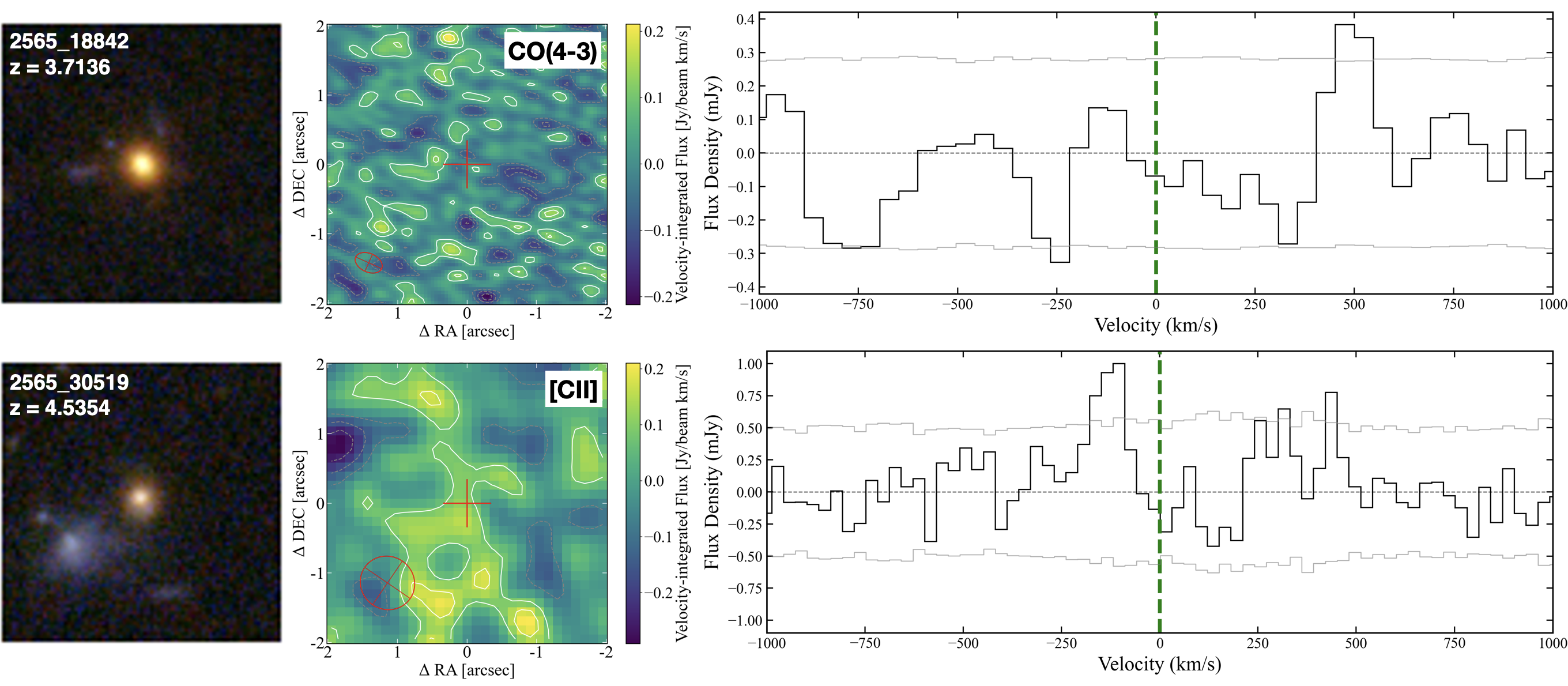} \hspace{-2mm}
\includegraphics[width=14.3cm]{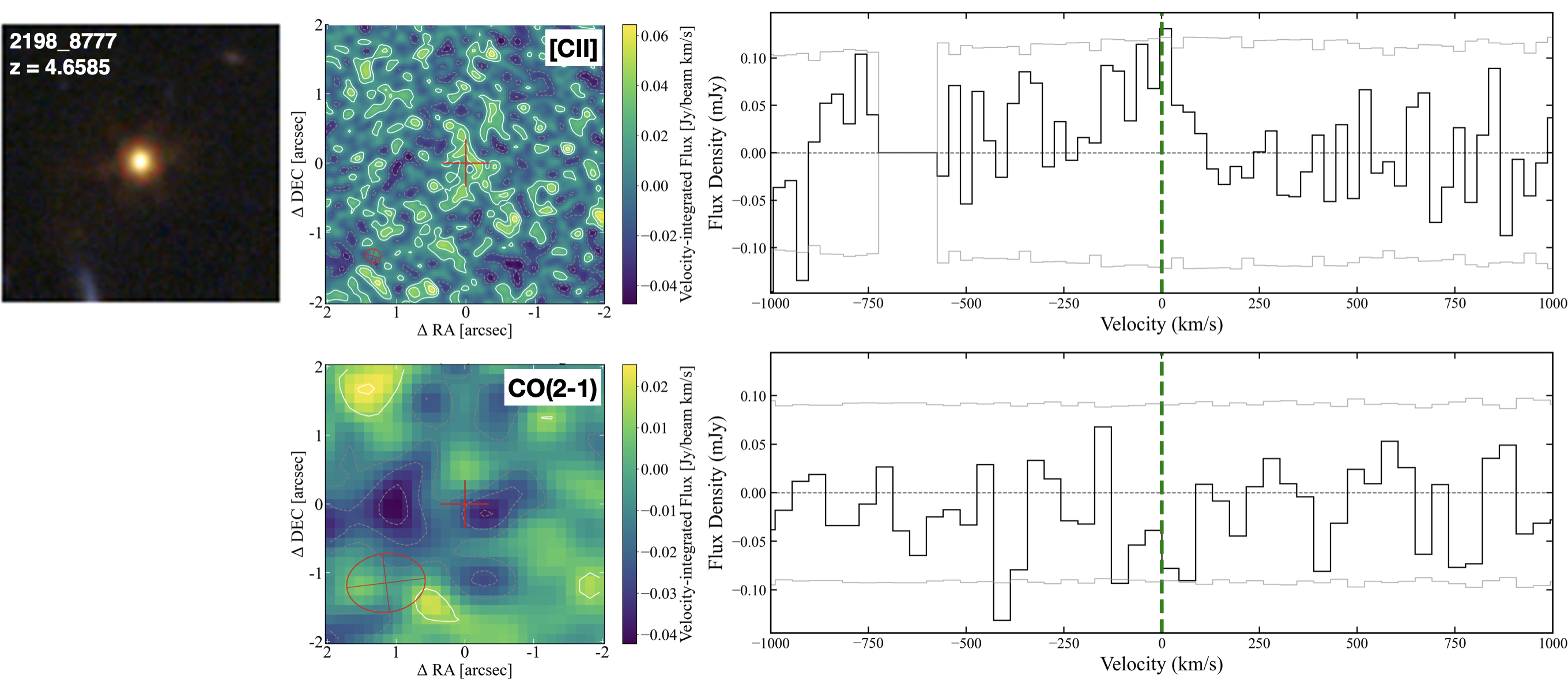} \hspace{-2mm}
\caption{\textbf{ALMA non-detections of \cii$\,158\,{\rm \mu m}$ and CO emission lines for the dormant galaxies.} \textit{Left}: $4^{\prime\prime}\times4^{\prime\prime}$ JWST RGB images (F115W in blue, F277W in green, and F444W in red). \textit{Middle}: ALMA moment-0 maps of the \cii or CO lines (assuming a line width of 250 km s$^{-1}$; $4^{\prime\prime}\times4^{\prime\prime}$). The contour levels start at 1$\sigma$ and increase in steps of $\pm1\sigma$, where positive and negative contours are solid and dashed, respectively. The synthesized beam is shown in the lower-left corner. \textit{Right}: ALMA 1D spectra binned to 30-50 km s$^{-1}$. The grey solid lines show the associated 1$\sigma$ uncertainty. The vertical green dashed lines mark the locations of expected \cii$\,158\,{\rm \mu m}$ or CO lines based on the JWST spectroscopic redshifts. The spectra are extracted at the source positions using beam-sized apertures, assuming that the emission line is unresolved. 
      }
\label{fig:almaDG1}
\end{figure*}

\begin{figure*}[h!]
\centering
\includegraphics[width=14.3cm]{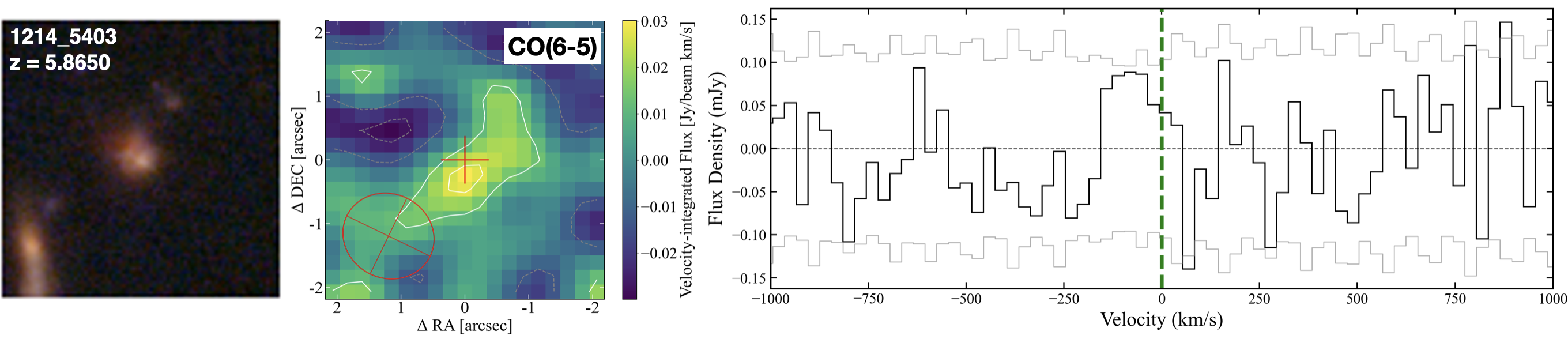} \hspace{-2mm}
\caption{Continued. 
}
\label{fig:almaDG2}
\end{figure*}

\end{appendix}

\end{document}